\def\update{\Large }

\documentclass{mn2e}
\usepackage{graphicx}
\usepackage{epsfig}
\usepackage{float}

\newcommand{\gae}
{\lower 2pt \hbox{$\, \buildrel {\scriptstyle >}\over {\scriptstyle \sim}\,$}}
\newcommand{\lae}
{\lower 2pt \hbox{$\, \buildrel {\scriptstyle <}\over {\scriptstyle \sim}\,$}}

\begin{document}

\title[Change in General Relativistic Precession Rates due to Lidov-Kozai oscillations in Solar System]
{Change in General Relativistic Precession Rates due to Lidov-Kozai oscillations in Solar System}

\author[A. Sekhar, D. J. Asher, S. C. Werner, J. Vaubaillon, G. Li]
{A. Sekhar$^{1,2*}$, D. J. Asher$^2$, S. C. Werner$^1$, J. Vaubaillon$^3$, G. Li$^4$\\
$^1$Centre for Earth Evolution and Dynamics, Faculty of Mathematics and Natural Sciences, University of Oslo, Blindern N-0315, Norway\\
 $^2$Armagh Observatory and Planetarium, College Hill, Armagh BT61\ 9DG, United Kingdom\\
 $^3$IMCCE, Observatoire de Paris, 77 Avenue Denfert Rochereau, F-75014 Paris, France\\
 $^4$Harvard-Smithsonian Center for Astrophysics, Cambridge, MA, USA \\
 $^*$E-mail: aswin.sekhar@geo.uio.no, asw@arm.ac.uk \\ }

\date{{\bf Accepted}: ; {\bf Received}: ; {\bf In Original Form}: ; {\bf MNRAS} \update }

\maketitle

\begin{abstract}
Both General Relativistic (GR) precession and the Lidov-Kozai mechanism,
separately, are known to play an important role in the orbital evolution of
solar system bodies. Previous works have studied these two mechanisms
independently in great detail. However, both these phenomena occurring at the
same time in real solar system bodies have rarely been explored. In this
work, we find a continuum connecting the GR precession dominant and
Lidov-Kozai like mechanism dominant regimes, i.e.\ an intermediate regime where
the competing effects of GR precession and Lidov-Kozai like oscillations co-exist
simultaneously. We find some real examples in the solar system in this
intermediate regime. Moreover we identify a rare example amongst them, comet
96P/Machholz 1, which shows significant changes in the rates of GR precession
(an order of magnitude higher than Mercury's GR precession rate) due to sungrazing
and sun colliding phases induced by Lidov-Kozai like oscillations. This comet's
combination of orbital elements and initial conditions (at the present epoch)
favour this measurable rapid change in GR precession (at some points peaking
up to 60 times Mercury's GR precession rate) along with prograde-retrograde
inclination flip (due to Lidov-Kozai like oscillations). Similar tests are
performed for hundreds of bodies lying in the moderately low perihelion
distance and moderately low semi-major axis phase space in the solar system,
the present lowest perihelion distance asteroid 322P/SOHO 1, and further
examples connected with 96P/Machholz 1 namely, the Marsden and Kracht
families of sungrazing comets plus low perihelion meteoroid streams like
Daytime Arietids (ARI) and Southern Delta Aquariids (SDA). 
\end{abstract} 


\section{Introduction} 

Two well known phenomena associated with low perihelion distance bodies in
orbital dynamics are general relativistic (GR) precession and Lidov-Kozai
oscillations. 

The accurate prediction of the perihelion shift of Mercury in accord with
real observations is one of the significant triumphs of the general theory of
relativity developed by Einstein (1915). Past works have looked into the GR
precession in perihelion in different types of solar system bodies like
planets (Weinberg 1972; Brumberg 1991; Quinn, Tremaine \& Duncan 1991; Iorio
2005), asteroids (Sitarski 1992), comets (Shahed-Saless \& Yeomans 1994) and
meteoroid streams (Fox, Williams \& Hughes 1982; Sekhar 2013; Galushina,
Ryabova \& Skripnichenko 2015). More recently some works have explored the
cases of GR precession in exoplanetary systems (Naoz et al.\ 2011, 2013; Li
et al.\ 2015).

A dynamical mechanism first found by Lidov (1962), and further studied
by Kozai (1962) who applied it to sun-Jupiter-asteroid 3-body systems,
explains the periodic exchange between eccentricities $e$ and inclinations
$i$ thereby increasing or decreasing the perihelion distance $q$ secularly in
the orbiting body. In its purest form the Lidov-Kozai mechanism involves
three bodies, namely a central body, test particle and perturber. In real
situations such as the solar system, where the perturber may be interior or
exterior (page 154, Morbidelli 2011) to the test particle and/or there may be
higher order multiplicity effects (due to perturbations from other giant
planets than Jupiter, lesser in strength compared to Jovian perturbations),
pure Lidov-Kozai still offers a convenient way to understand the dynamical
behaviour. We shall use the term Lidov-Kozai like mechanism to cover these
situations.

This mechanism has been related to the rapid change in orbits of artificial
satellites (Lidov 1962) around the Earth. Past works have shown that the
Lidov-Kozai like mechanism can lead to a flip in orbits i.e.\
inclinations switching from prograde to retrograde or vice-versa (Naoz
et al.\ 2011; Lithwick \& Naoz 2011; Naoz et al.\ 2013; Li et al.\ 2014b)
during the body's secular evolution. Naoz et al. (2011) discussed the
possibility of orbital flips in hierarchical triple body systems for the
first time in the context of exoplanet systems. Bailey, Chambers \& Hahn
(1992) found that Lidov-Kozai (here it is not pure Lidov-Kozai mechanism
in strict terms but higher order multiplicity effects) is an efficient way
by which asteroidal and cometary orbits could land up in sunskirting or
sungrazing orbits. The mechanism is known to have an important role in the
long term evolution of different classes of small bodies (Vaubaillon, Lamy \&
Jorda 2006; Granvik, Vaubaillon \& Jedicke 2012) in the context of impact
studies (Werner \& Ivanov 2015) due to the Lidov-Kozai like cycles in
orbital elements which lead to complications in using analytical and
numerical techniques (Manley, Migliorini \& Bailey 1998) to compute impact
probabilities from small bodies on planets. More recently there are examples
found in the exoplanetary systems (Naoz et al.\ 2013; Li et al.\ 2014a; Naoz
2016) which undergo Lidov-Kozai like oscillations around the central
body thus showing the generality of this phenomenon in any suitable dynamical
system.

Of the past studies mentioned above, the ones that pertain to the solar
system have generally related to either of the two phenomena separately, not
GR and Lidov-Kozai like effects happening at the same time. Moreover
previous works in exoplanet systems have indicated that GR precession can
suppress Kozai like oscillations in some cases (Naoz et al. 2013)
depending on the orbital elements phase space. Hence it has been shown that
one phenomenon can compete or dominate over another in exoplanet
systems. Again, this particular idea has not been well explored in the
context of real solar system bodies. Naoz et al. (2013) showed that inclusion
of GR precession can lead to re-triggering eccentricity excitations and
orbital flips thereby pointing to the situation of both GR precession and
Kozai effect co-existing at the same time in exoplanet systems.

In this work, we find real examples of solar system bodies where
significant (to be quantified later) GR precession and Lidov-Kozai like
behaviour can co-exist and complement each other, forming a continuum between
the two regimes where the respective separate effects dominate. The
Lidov-Kozai like mechanism leads to secular lowering of $q$ which in turn
leads to a huge increase in GR precession of argument of pericentre
$\omega$. This in turn gives feedback to the Lidov-Kozai like mechanism
as the $e$, $i$ and $\omega$ cycles are closely correlated.

A real solar system body, 96P/Machholz 1, exhibiting these trends by
combining the two dynamical effects during present times is identified out of
the hundreds of small bodies lying in the moderately low $q$ and moderately
low $a$ (semi-major axis) phase space. This particular effect would be more
pronounced especially when a body gradually evolves into a sungrazing and sun
colliding orbit. The dynamical evolution becomes even more interesting when
inclination flips occur, the secular Lidov-Kozai like reduction in $q$
applying in both prograde and retrograde cases. 

For all calculations in this paper, Newtonian n-body integrations were
done using Chambers' (1999) MERCURY package incorporating the RADAU (Everhart
1985) and MVS (Mixed-Variable Symplectic: Wisdom, Holman \& Touma 1996)
algorithms. For GR cases, MERCURY integrations included an additional
sub-routine (provided by G. Li) incorporating the GR corrections (Benitez \&
Gallardo 2008) in the n-body code; the MVS algorithm was used. The RADAU
accuracy parameter was set to 10$^{-16}$ and the MVS time step was chosen as
one day. Each asteroid, comet and meteoroid particle was treated as a test
particle (i.e.\ zero mass) and integrated in the combined presence of the sun
and eight planets Mercury to Neptune. Non-gravitational forces were not
considered in comets, nor radiative forces for asteroids and meteoroid
streams, so that we can distinguish GR active and Kozai like dynamics active
spaces without additional complicating effects. In the GR-included
integrations, GR effects were taken into account for the evolution of all the
big bodies (i.e.\ planets) and small bodies involved. In the Newtonian
integrations, GR did not act on either big or small bodies. For all
integrations the initial orbital elements (taken from JPL-Horizons: Giorgini et al.\ 1996) 
and initial epoch (JD 2451000.5) of the eight planets remained the
same, for uniformity in comparisons. Hence every GR case presented here, and
likewise every Newtonian case, is based on an identical solar system
model. Cross-checks between RADAU and MVS Newtonian integrations confirmed
that either integrator could reproduce the same orbital evolution. Searches
for small bodies with various ranges of orbital elements used IAU-MPC (Minor
Planet Center) and initial elements of specific bodies to be integrated were
the latest epoch available from JPL-Horizons.

\begin{table*}
\caption{Initial conditions for the bodies presented as plots in this
work. All taken from JPL-Horizons except for meteoroid stream Daytime
Arietids (ARI) from IAU-Meteor Data Center.}
\begin{tabular}{ccccccccc}
 & Epoch JDT & $a$ & $e$ & $i$ & $\omega$ & $\Omega$ & $M$ & Fig. \\
C/1932 G1 & 2426760.5 & 45.00523804 & 0.972127 & 74.277600 & 303.5157 & 213.4835 & 359.980648 & \ref{F1932G1} \\
2011 CL50 & 2455606.5 &0.88303406 & 0.147614 &0.173164 &285.543879 &22.791201 &201.817484 & \ref{FCL50} \\
96P & 2456541.5 &3.03393972 &0.959211 &58.312214&14.757748&94.323236 &77.992760 & \ref{ei96P} \\
ARI & 2457546.5 &2.67 &0.974 &27.7 &28.7 &79.1 &0.0 & \ref{ARI} \\
322P & 2457179.5 &2.51626146 &0.978676 &12.589235 &49.049476 &359.524487 &337.760658 & \ref{322P} \\
2008 KP & 2455509.5 &1.10063074 &0.789847 &59.835824 &344.973090 &62.464197 &293.337766 & \ref{2008KP} \\
Mercury &2451000.5 &0.38709895 &0.205620 &7.005045 &29.121849 &48.332314 &106.519879 & \ref{PrecMercury} \\
\end{tabular}
\label{tab-els}
\end{table*}

\begin{figure}
(a)\\[-\baselineskip]
\includegraphics[width=\columnwidth]{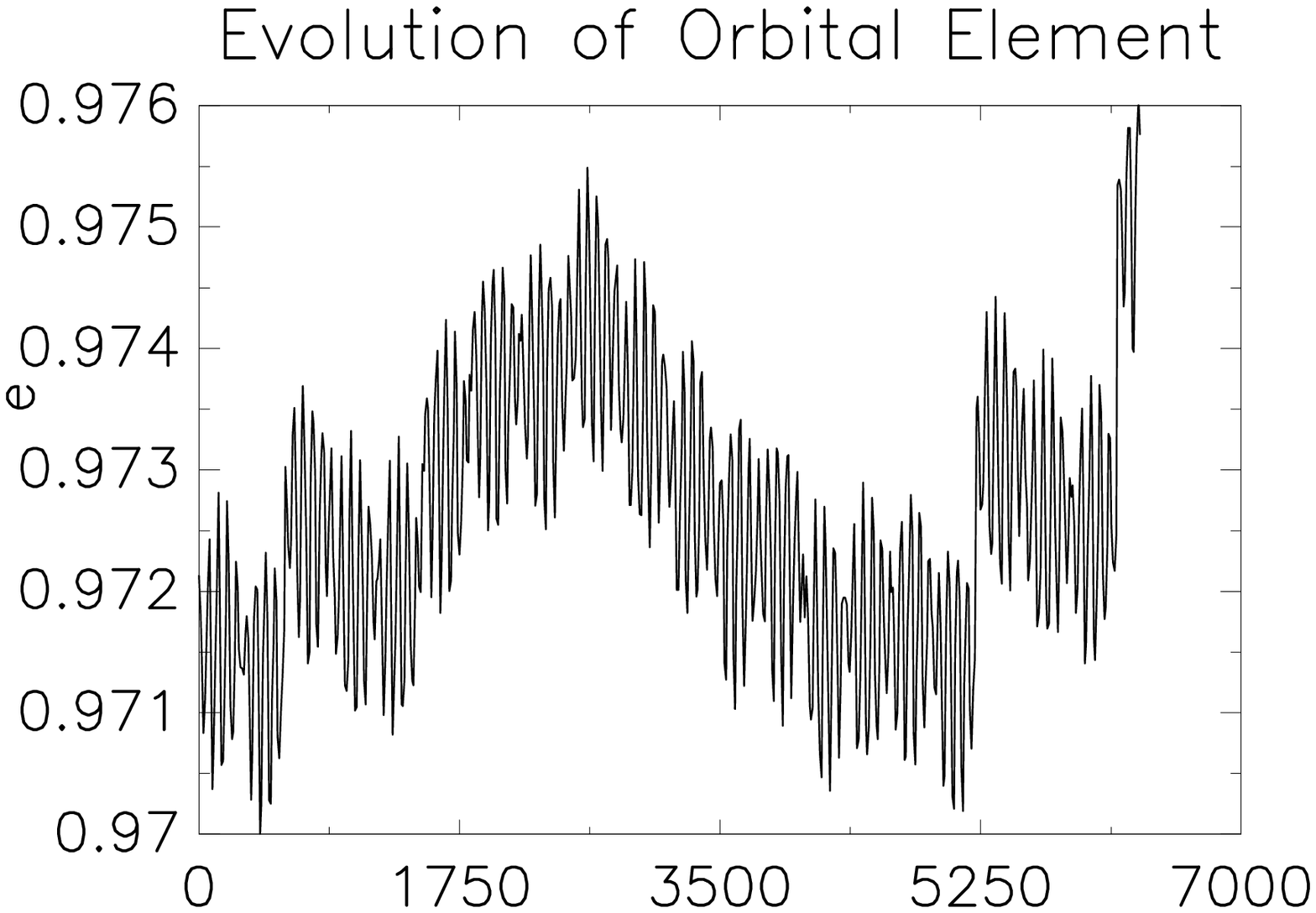}
\\[-4mm]
(b)\\[-\baselineskip]
\includegraphics[width=\columnwidth]{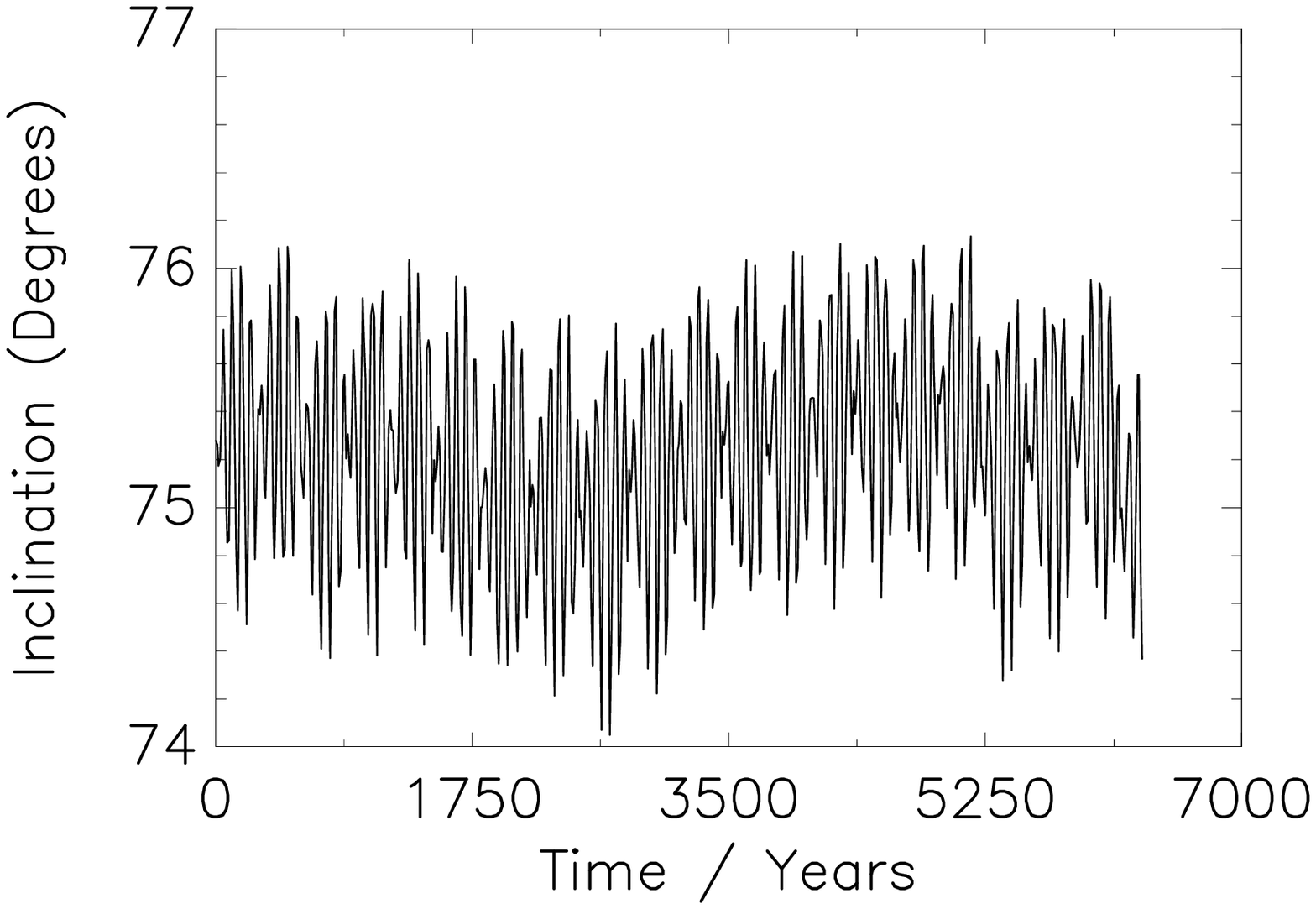}
\\[-4mm]
\caption{Orbital evolution of (a) eccentricity (b) inclination
of C/1932 G1 (Houghton-Ensor) forward for 7 kyr from present. This body gets
ejected out of solar system in about 6.3 kyr in our simulations. }
\label{F1932G1}
\end{figure} 

\section{Lidov-Kozai like mechanism dominant cases}
\label{KMvsGR}

Long term Lidov-Kozai like oscillations can be found in different examples of
sungrazing comets (Bailey, Chambers \& Hahn 1992), bringing these comets
closer and closer to the sun after every perihelion passage and $q$ being
lowered secularly. Although the perihelion distances stay in the range of
GR-active space and can, in principle, increase the GR precession rates (due
to decrease in $q$), the orbital periods $P$ (and $a$) mostly stay way beyond
the significant GR-active space. For GR effects happening near the perihelion
passage trajectory to accumulate efficiently over time, the body needs a
relatively small orbital period so that perihelion passages happen
frequently. Hence for most known bodies this GR precession increase is
insignificant and difficult to separate and confirm from observations. 
In this section, we talk about single orbit timescales of the order of $\sim
10^{2}-10^{6}$ yr, so that even if GR precession per orbit is significant, the
accumulated effect over these timeframes is low.

In this work we quantify GR
precession as significant if it is comparable to that of Earth. For Earth's
present orbit $\Delta \omega \sim$ 3.8 arc seconds per century or
$\Delta \omega \sim$ 0.038 arc seconds per perihelion passage (table 8.3,
page 198, Weinberg 1972) which has been verified and confirmed by
observations. 

GR precession in $\omega$ can be computed using this closed form expression
(page 197, Weinberg 1972):
\begin{equation}
\Delta \omega = 6 \pi GM/a(1-e^{2}) = 6 \pi GM/q(1+e)
\end{equation}
in radians/revolution. It should be noted that the direction of the
precession of pericentre is always in the same direction of the motion of the
orbiting body (page 197, Weinberg 1972). 

In the case of Halley-type comets (with relatively low $q$ $\leq$ perihelion
distance of Mercury), the periods (order of $\sim 10^2$ yr) are high enough
that it is practically difficult to distinguish GR precession effects (from
other effects or changes in orbits) after each perihelion passage and
moreover, the number of perihelion passages to accumulate GR precession
(during every near $q$ passage) is small per unit of time. However
Kozai like oscillations can be easily distinguished in these bodies during their
long term evolution and the Kozai like mechanism stays dominant (over GR
precession) in these bodies. This point about insignificant GR precession
accumulation applies to other minor bodies such as Edgeworth-Kuiper Objects
which have relatively high $P \ga 10^2$ yr. 

Figure \ref{F1932G1} shows the $e$ and $i$ evolution of C/1932 G1 (a representative
Halley-type low-$q$ comet). Kozai like oscillations can be seen in plots (a)
and (b). Tests were repeated for Halley-type comets taken from the cometary
catalogue compiled by Marsden \& Williams (2008) and they show insignificant
GR precession as expected, Kozai like mechanism dominant over GR precession being
the typical behaviour because of the higher $P$.

\begin{table}
\caption{GR precession in argument of pericentre $\Delta \omega$ per
perihelion passage for examples of highly eccentric sungrazing comets
(orbital elements taken from sungrazing comets section of Marsden \& Williams
2008). For comparison, the value for Mercury is 0.103 arc seconds/revolution.
}
\vspace{0mm}
\begin{center}
\begin{tabular}{|l|l|l|l|l|r|r|} \hline
Body & $q$ & $e$ &$\Delta \omega$ & $P$   \\
& (AU)           &      & (arc seconds      & (yr)  \\
&                    &      & per $q$ passage) &         \\
\hline \hline
C/1979 Q1 & 0.005                     & 1                   & 4.0 & NA     \\

C/1965 S1-A & 0.008                     & 1                & 2.5 &  900  \\

C/2008 E7  & 0.055                     &1                    &0.35       &NA    \\

C/1997 H2  & 0.136                    & 1                    &0.14     & NA  \\
\hline
\end{tabular}
\end{center}
\label{tab-sungraz}
\end{table}

Long period sungrazers like the Kreutz family of comets (Kreutz 1888) and Meyer group
come incredibly close to the sun. In these cases, although the GR perihelion precession per
single perihelion passage is at the highest (see Table \ref{tab-sungraz}),
$P$ is of the range $\sim 10^{3} - 10^{6}$ yr (mostly members of the Kreutz family) and hence GR precession
accumulating over time is out of the question for the timeframes we discuss
in this work. Moreover because of the perturbations from galactic tides and
passing stars when these bodies reach near aphelion, it would be difficult to
imagine any of these bodies showing a consistent or periodic pattern of GR
precession accumulation in comparable magnitude over long time scales. 

The examples presented in Table \ref{tab-sungraz} are representatives showing
low, intermediate and high $q$ amongst the sungrazing candidates ($e \sim 1$)
in the Marsden \& Williams (2008) comet catalogue. Table \ref{tab-sungraz}
shows that lowest $q$ candidate can have GR precession per revolution as much
as 40 times that of Mercury's GR precession per revolution but because of
large $P$ of typical sungrazing comets, this precession accumulating
efficiently over time is unrealistic for future journeys into the inner solar
system. Moreover some extreme sungrazers do not survive for another
perihelion passage due to their disruption near the sun or collision with the sun. 

However in some particular cases in the solar system, the GR precession can
be significant and measurable, while Lidov-Kozai like cycles still survive during
the same time (Section \ref{GRandKM}).

\section{GR precession dominant cases}
\label{GRvsKM} 
GR precession in pericentre was first studied for the classic example of
Mercury's orbit. More recently there are works (Kerr 1992; Bou\'e, Laskar \&
Farago 2012; Lithwick \& Wu 2014) which looked into the long term dynamical
behaviour of Mercury and its stability within the solar system. These latest
works have indicated that GR precession along with secular resonances are the
dominating effects in Mercury's long term evolution ruling out an important
role from any Kozai like mechanism. In this section, we talk about GR
accumulation with single orbit timescales of the order of $\leq$ 1 yr. In
these cases, the GR effect is accumulated slowly and steadily over time.

\begin{table}
\caption{GR precession in argument of pericentre $\Delta \omega$ per century
for some examples of low $q$, low $a$ and low $i$ bodies in solar system. For
comparison, the value for Mercury is 43 arc seconds/century.} 
\vspace{0mm}
\begin{center}
\begin{tabular}{|l|l|l|l|l|r|r|} \hline
Body & $q$ & $a$ &$\Delta \omega$ & $\Delta \omega$   \\
& (AU)           &(AU)      & (arcseconds      & (arcseconds   \\
&                    &      &      per century)         &     per century)       \\
&                    &      & (analytic)   &  (integrations)        \\
\hline \hline
2015 KE & 0.85                   & 0.97                  &4.2  &4.6    \\ 

2013 BS45 & 0.91                   & 0.99                  &4.0  & 4.5    \\ 

2010 VQ & 0.69                    &0.86               &5.8     & 6.2  \\ 

2011 CL50  & 0.76                    &0.89          & 5.3      &5.9    \\ 
\hline
\end{tabular}
\end{center}
\label{tab-low-a}
\end{table}

\begin{figure}
(a)\\[-\baselineskip]
\includegraphics[width=\columnwidth]{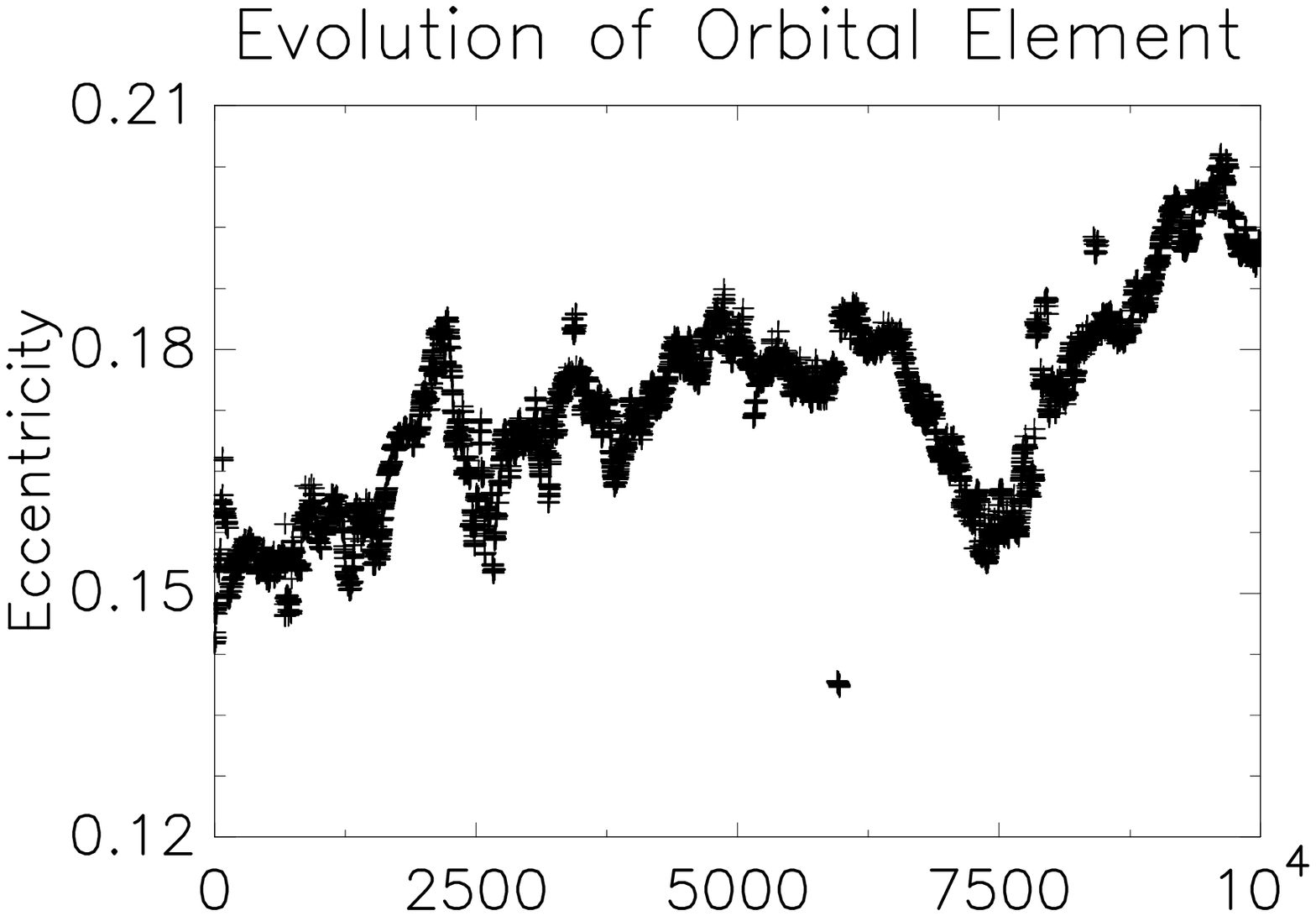}
\\[-5mm]
(b)\\[-\baselineskip]
\includegraphics[width=\columnwidth]{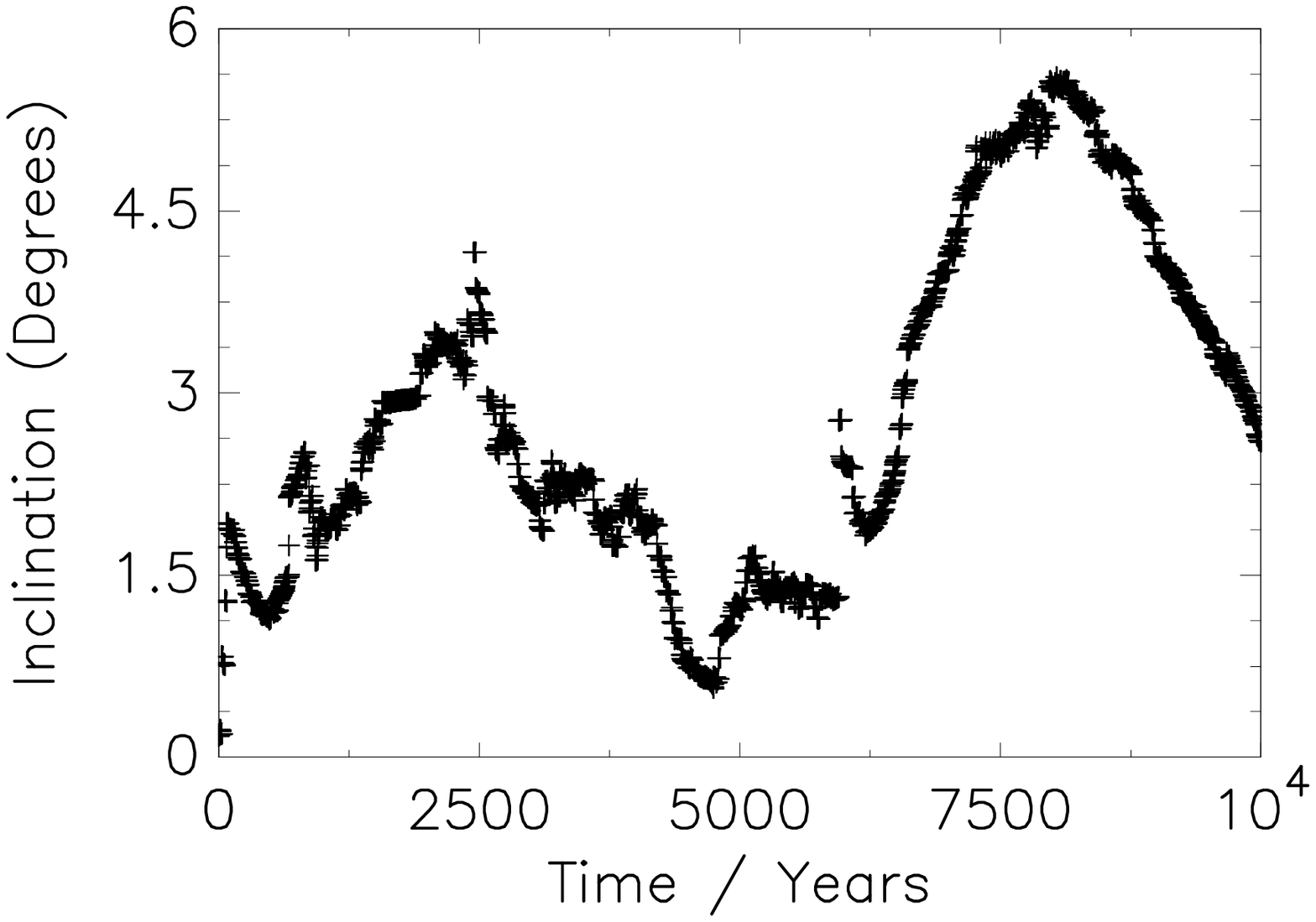}
\caption{Evolution of $e$ and $i$ of 2011 CL50 for 10 kyr into future. }
\label{FCL50}
\end{figure}

Here we look at the evolution of bodies within $e \leq$ 0.2 ($\sim$
eccentricity of Mercury), $i \leq$ 7.0 ($\sim$ inclination of Mercury) and
$a \leq 1.0$ au ($\sim$ semi-major axis of Earth). This range enlists 115
objects in the IAU-MPC database. Numerical integrations of these bodies show
measurable GR precession (which can be tested using comparison of analytical
calculations and numerical integrations; see Table \ref{tab-low-a}) and
Lidov-Kozai like oscillations at some point in the future if the
evolution is followed for a few $10^5$ yr. However all the subset of these
bodies with $i \leq$ 3\degr\ (61 objects) do not show (see the example in
Figure \ref{FCL50}) consistent and long term Lidov-Kozai like
oscillations for the near future ($\sim 10^3$ yr). The bodies shown in
Table \ref{tab-low-a} (the last of which is plotted in
Figure \ref{FCL50}) are representative of these 61 objects. Because both $e$
and $i$ are low at the same time in these cases compared to bodies
discussed in Section \ref{KMvsGR}, the Lidov-Kozai like oscillations do
not dominate over other effects like GR precession or secular
resonances. It should be noted that orbital timescales depend only on
semi-major axis and eccentricity here. But inclination is mentioned and
discussed because anti-correlation of $e$ and $i$ is a crucial signature of
Kozai like oscillation and the nature of inclination evolution helps to
identify Kozai like behaviour in the solar system.

The $e$ and $i$ evolution of 2011 CL50 (Figure \ref{FCL50}) shows no consistent
Kozai like oscillations whereas the GR precession is significant
(Table \ref{tab-low-a}). This trend is typical for other low $q$, low $a$ and
low $i$ bodies from the 61 bodies enlisted and discussed in this section. The
cases showing absence of consistent Lidov-Kozai like oscillations (or negative
results) are presented here to highlight the fact that there are areas in the
solar system where significant GR precession dominates over Kozai like mechanism
(similar to the exoplanetary cases presented in Naoz et al. 2013), in
contrast to examples in Section \ref{KMvsGR}. The overlap of these phenomena
(GR precession and Kozai like oscillations), complementing each other thereby
leading to unique dynamical behaviour, is discussed in detail in Section
\ref{GRandKM}. 

\section{Co-existence of GR precession and Lidov-Kozai oscillations} 
\label{GRandKM} 
In this section, we consider Lidov-Kozai like cycle timescales of the
order of some kyr which contributes to GR enhancement due to sungrazing phases
induced by a Kozai like mechanism. 
We consider bodies with $q \leq$ 0.3 au ($\sim$ perihelion distance of
Mercury) and $a \leq 4$ au ($\sim$ 10 times the semi-major axis of Mercury),
thus single orbit timescales of the order of $\leq$ 8 years. 
This condition enlists 244 objects from the IAU-MPC database. The same
condition was applied to the list of established meteor showers from the
IAU-MDC (Meteor Data Center) database. Some representative bodies with relatively high GR
precession rates from these lists and their GR precession calculated using
analytical formulae and numerical integrations are shown in
Table \ref{tab-var}.

\begin{figure}
(a)\\[-\baselineskip]
\includegraphics[width=\columnwidth]{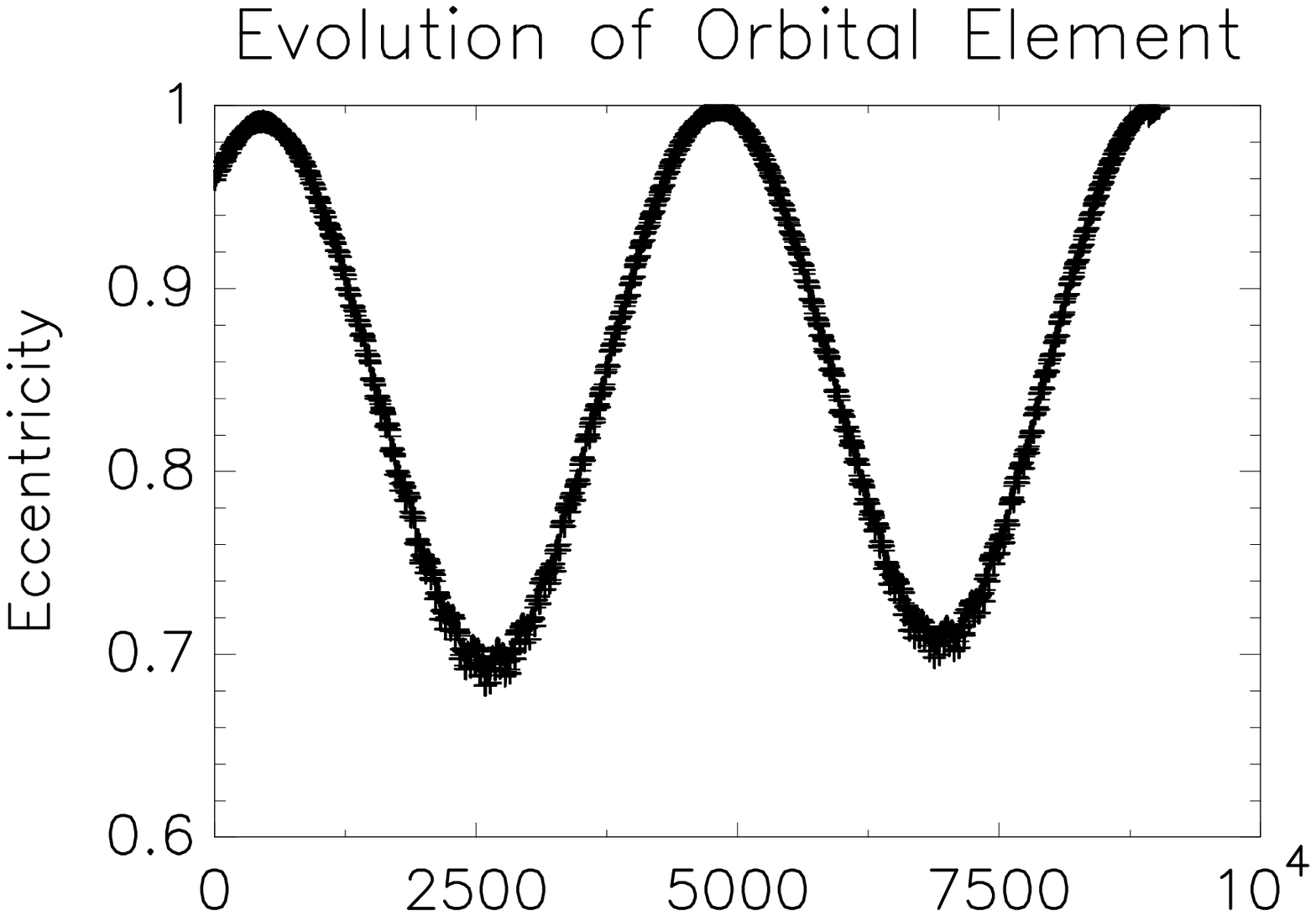}
\\[-5mm]
(b)\\[-\baselineskip]
\includegraphics[width=\columnwidth]{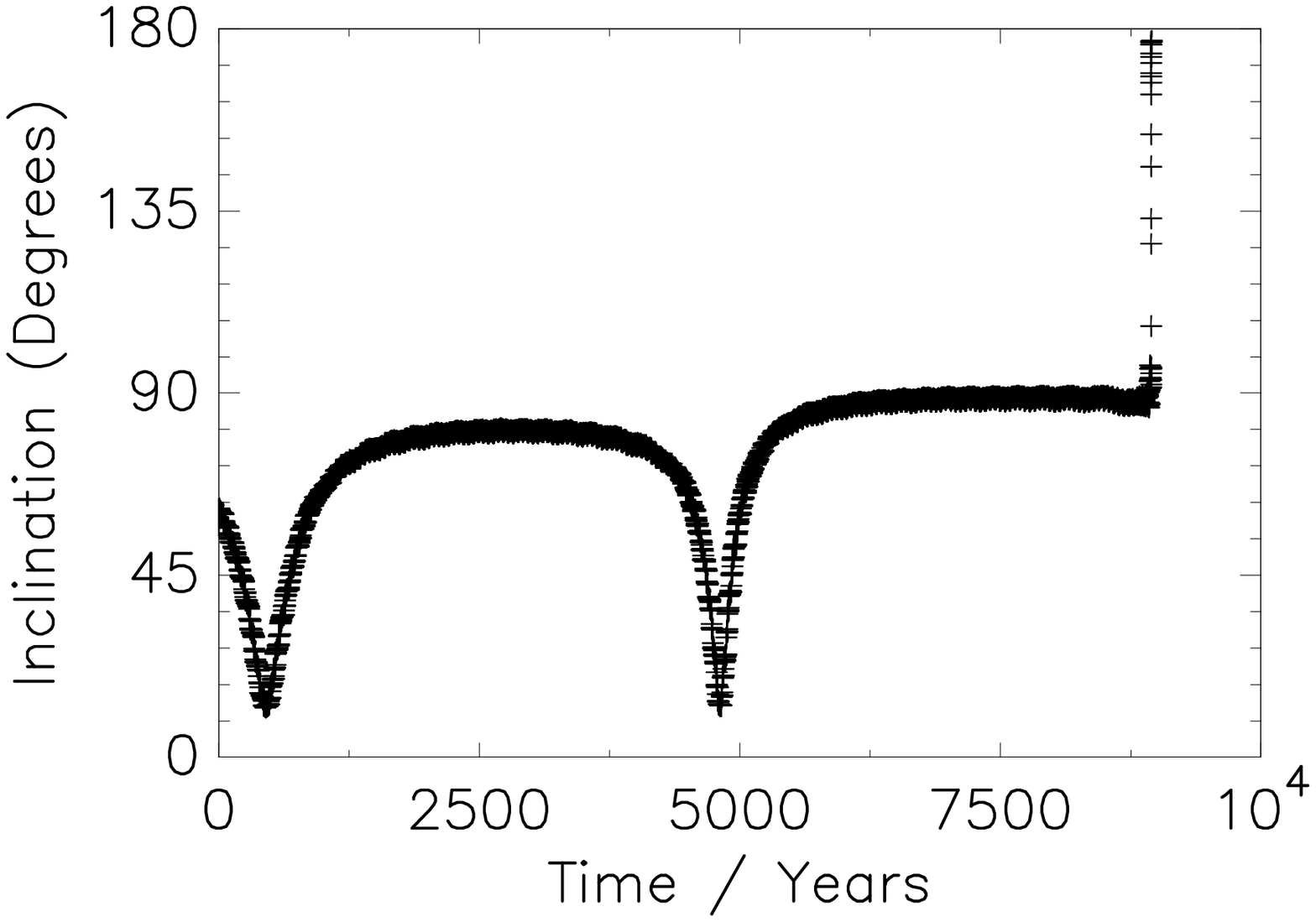}
\\[-2mm]
(c)\\[-\baselineskip]
\includegraphics[width=\columnwidth]{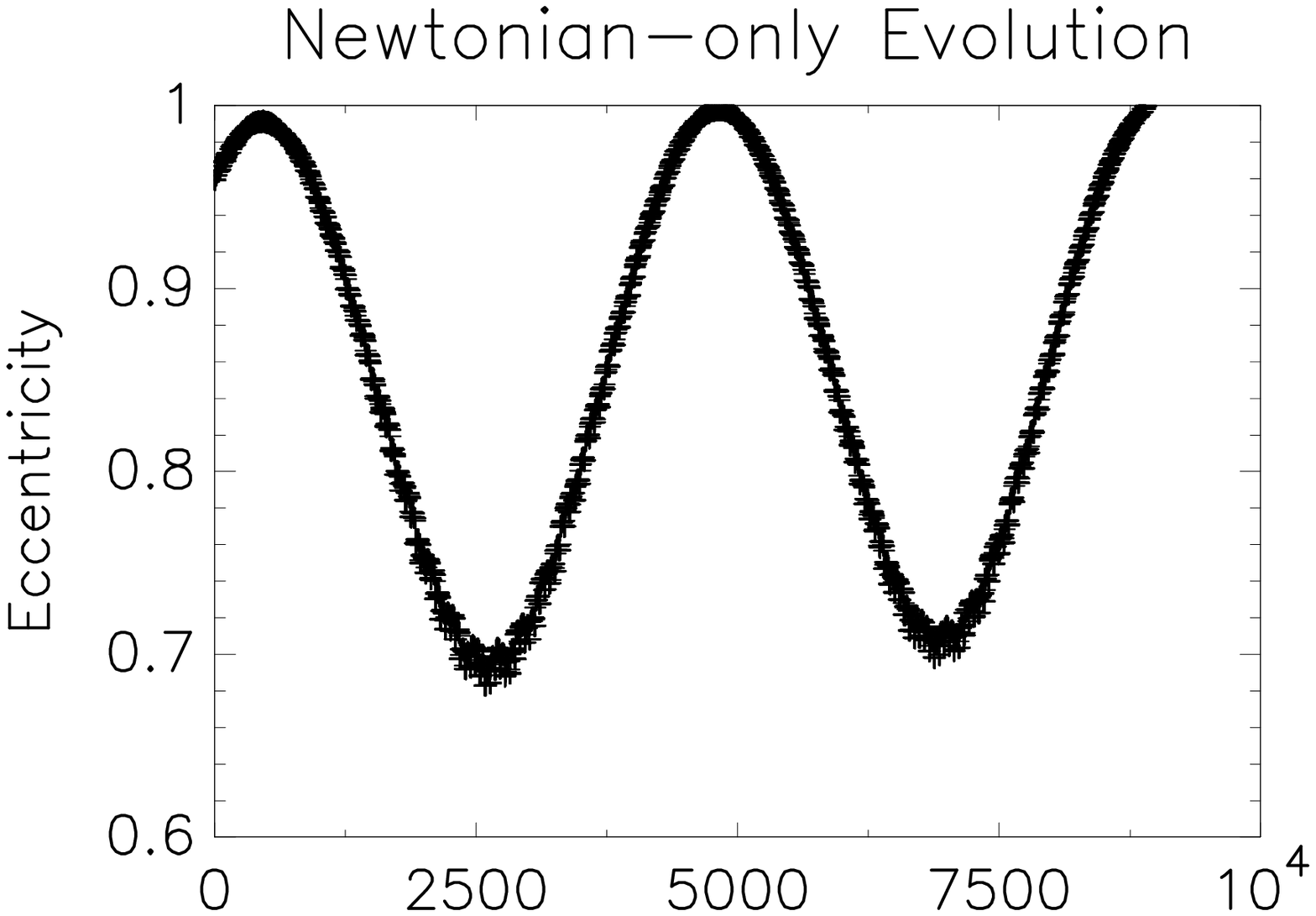}
\\[-5mm]
(d)\\[-\baselineskip]
\includegraphics[width=\columnwidth]{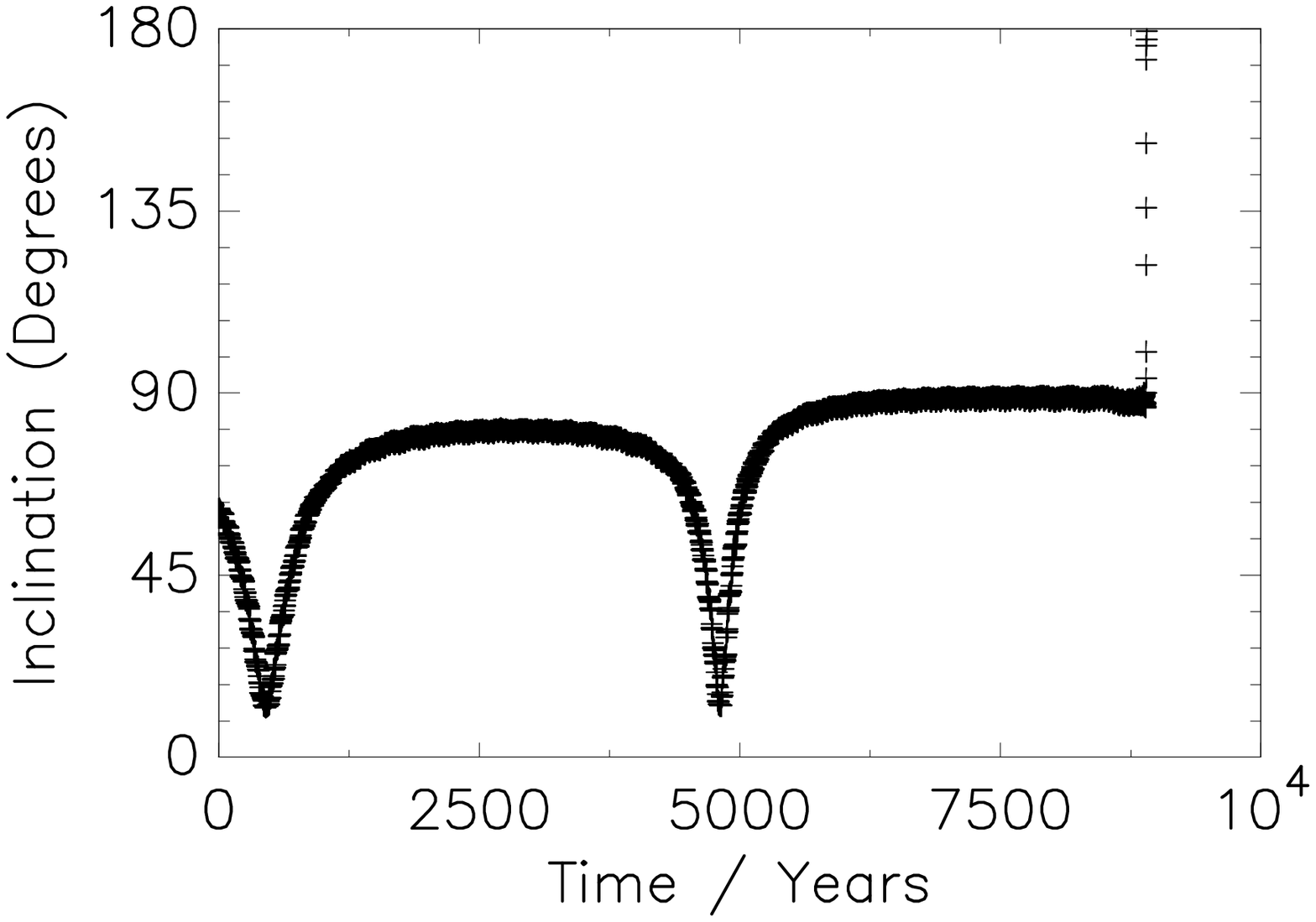}
\caption{Orbital evolution of (a,c) eccentricity (b,d) inclination of
96P for 9 kyr for GR-included and Newtonian-only cases respectively. Kozai like oscillations in $e$ and $i$
can be seen.}
\label{ei96P}
\end{figure}

We were interested to identify bodies evolving in the near future ($\sim$
thousands of years) into rapid sungrazing and sun colliding phases and
undergoing inclination flips, due to Lidov-Kozai like oscillations and being GR
active at the same time. Of all the bodies we checked from the IAU-MPC, and
Marsden plus Kracht families from the comet catalogue (Marsden \& Williams
2008), 96P/Machholz 1 stands out because it shows all these trends in the
near future. The uniqueness of 96P has been reported before in different
contexts, its dynamical behaviour having octuple crossing possibilities
(Babadzhanov \& Obrubov 1987, 1992a, 1992b, 1992c, similar to the case of
another periodic comet Machholz, 141P, discussed in Asher \& Steel 1996). The
linkage of the orbit of 96P with orbits of Extreme Trans Neptunian Objects
(ETNOs) has been explored by de la Fuente Marcos, de la Fuente Marcos \&
Aarseth (2015). 96P has been linked with two families of sungrazing comets
and two meteoroid streams.  Because of its previously established connection
(Ohtsuka, Nakano \& Yoshikawa 2003, Sekanina \& Chodas 2005) with the Marsden
and Kracht sungrazers and low $q$ meteoroid streams like Daytime Arietids
(ARI) and Southern Delta Aquariids (SDA), our tests were repeated on all
these related objects as well. 

Figures \ref{ei96P}a, \ref{ei96P}b and \ref{ei96P}c, \ref{ei96P}d show the near future $e$ and $i$ evolution of 96P for GR-included and Newtonian-only cases respectively. The test particle undergoes Kozai like oscillation (cf. Abedin et al. 2017) and near the
final phase of about 120 yr, inclination flip occurs from prograde to
retrograde. By about 9 kyr the particle falls into the sun due to rapid
decrease in $q$ due to Lidov-Kozai like mechanism and eventually reaches
near-ecliptic inclination $i \sim$ 180\degr\ close to the timeframe of
collision with the sun. The general behaviour evident in Figure \ref{ei96P} of GR-included and Newtonian-only dynamical evolution being very similar is confirmed by integration of clones (see later, where the small but significant difference will
also be discussed). This trend of general dynamical behaviour in $e$ and $i$
being nearly identical between Newtonian-only and GR-included cases holds true
for other bodies discussed in this work. 

\begin{table}
\caption{GR precession in argument of pericentre $\Delta \omega$ per century.
For comparison, the GR value for Mercury is 43 arc seconds/century. The
orbital elements for meteoroid streams and other minor bodies are taken from
IAU-Meteor Data Center and JPL-Horizons respectively.
Amongst the bodies listed here, the long term orbital evolutions for
96P, Daytime Arietids and 322P are shown in Figures \ref{ei96P}, \ref{ARI}
and \ref{322P} respectively. The short term evolutions of GR precession rates
for 96P are shown in Figures \ref{PrecCentury}, \ref{PrecRevolution}
and \ref{PrecShortTerm}.}
\vspace{0mm}
\begin{center}
\begin{tabular}{@{}l|l|l|l|l@{}} \hline
Body & $q$ & $a$ &$\Delta \omega$ & $\Delta \omega$   \\
& (au)           & (au)     & (arcseconds      & (arcseconds   \\
&                    &      &      per century)         &     per century)       \\
&                    &      & (analytic)   &  (integrations)        \\
\hline \hline
96P/Machholz 1  & 0.124                     & 3.035                  & 3.0 &3.7     \\

322P/SOHO  & 0.054                     & 2.516                  & 9.0 &9.6     \\

1566 Icarus    & 0.187                     & 1.078                  & 10.0 &10.6     \\

3200 Phaethon  & 0.140                   &1.271               &10.1     & 10.7  \\

Geminids            &0.2                      &1.3             & 9.3 &9.9   \\

Daytime Arietids   & 0.08                 &2.7         &5.7       &6.5     \\

S. Delta Aquariids    &0.07      &0.98         &6.7 & 7.3     \\
\hline
\end{tabular}
\end{center}
\label{tab-var}
\end{table}

\begin{figure}
(a)\\[-\baselineskip]
\includegraphics[width=\columnwidth]{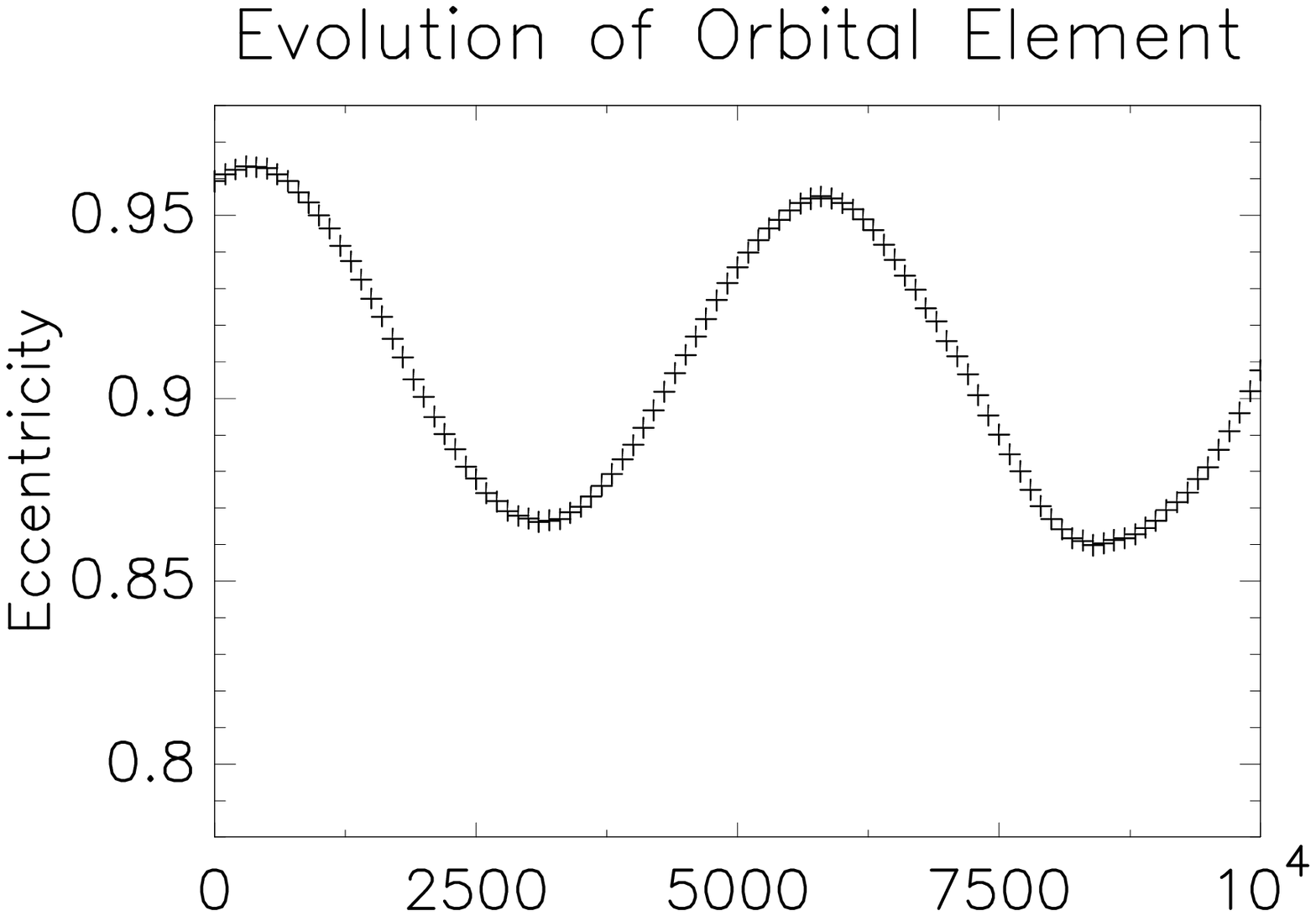}
\\[-4mm]
(b)\\[-\baselineskip]
\includegraphics[width=\columnwidth]{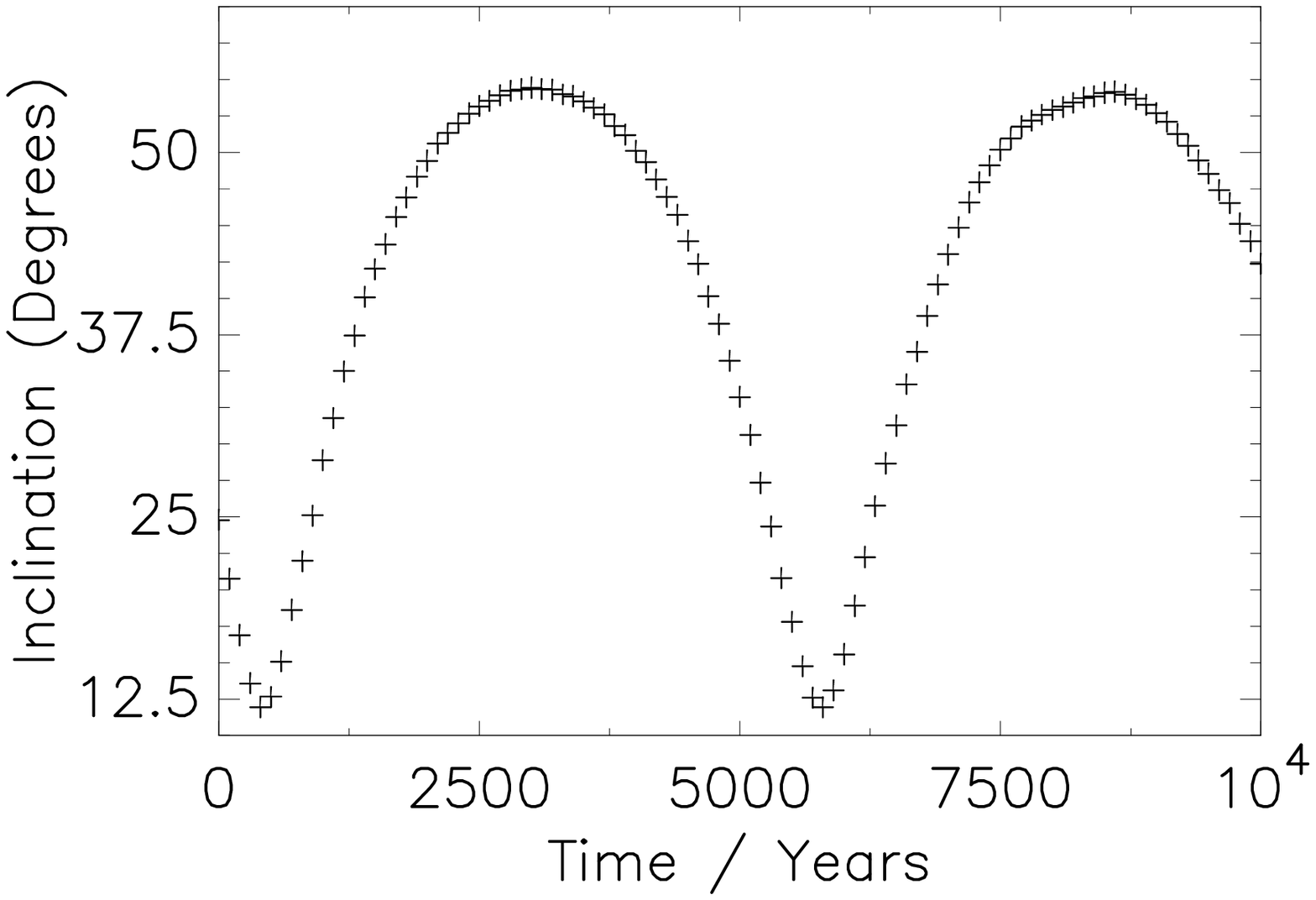}
\caption{Evolution in $e$ and $i$ of Daytime Arietids (ARI) 10 kyr into
future. Initial conditions are given in Table \ref{tab-els} which were
taken from IAU-MDC referring to observations reported in Jenniskens et
al. 2016.} 
\label{ARI}
\end{figure}

Figure \ref{ARI} shows the future orbital evolution of ARI.  Kozai like oscillations are
apparent and the same pattern applies to SDA (not shown here).

Among the highest $i$ set from the 244 objects mentioned at the start of this
section are the lowest perihelion distance asteroid 322P (Figure 5),
designated a comet but whose characteristics such as composition, density and
lack of activity point to asteroidal origin (Knight et al.\ 2016), and high
$i$ asteroids 2008 KP (Figure \ref{2008KP}), 385402, 333889 and 2010 KY127. All these
undergo Kozai like oscillations and GR precession at the same time. However we are
unable to find any inclination flips or sun colliding phase in these bodies
in the near future other than the unique example of 96P. Hence 96P stands out
in terms of its dynamical behaviour compared to all other bodies discussed in
this section. 

\begin{figure}
(a)\\[-\baselineskip]
\includegraphics[width=\columnwidth]{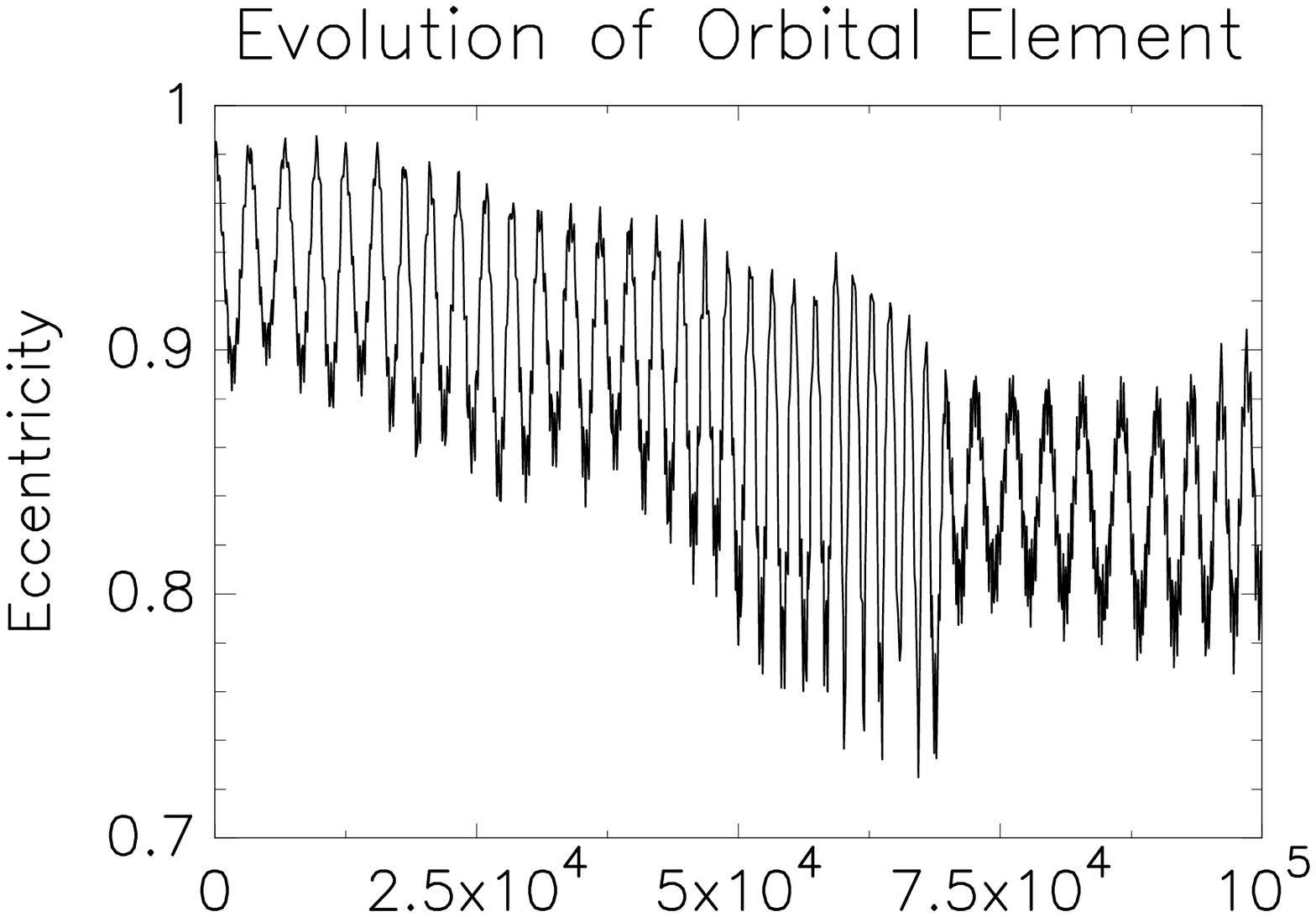}
\\[-4mm]
(b)\\[-\baselineskip]
\includegraphics[width=\columnwidth]{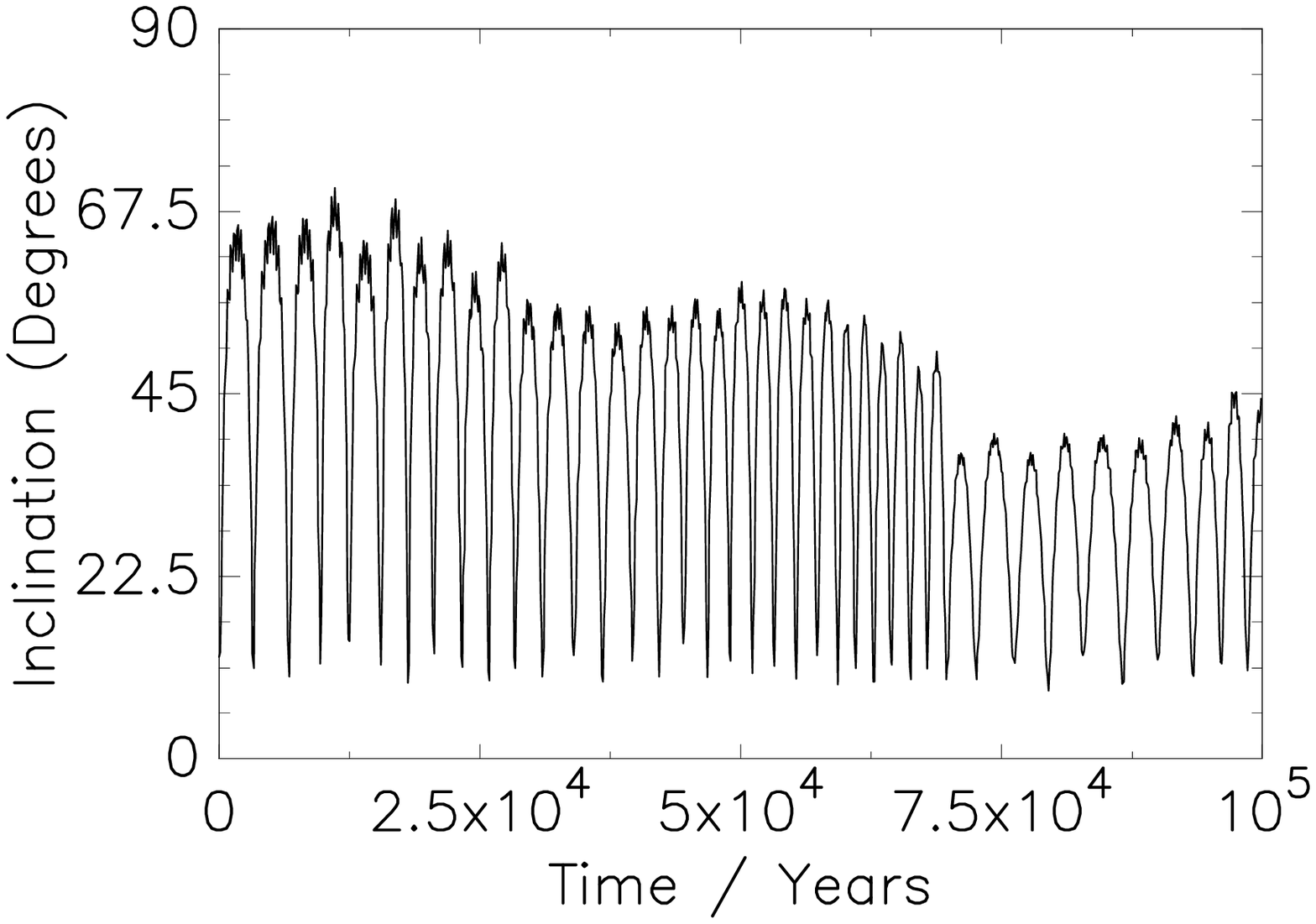}
\caption{Orbital evolution of (a) eccentricity (b) inclination
in lowest perihelion distance asteroid 322P. } 
\label{322P}
\end{figure}

\begin{figure}
(a)\\[-\baselineskip]
\includegraphics[width=\columnwidth]{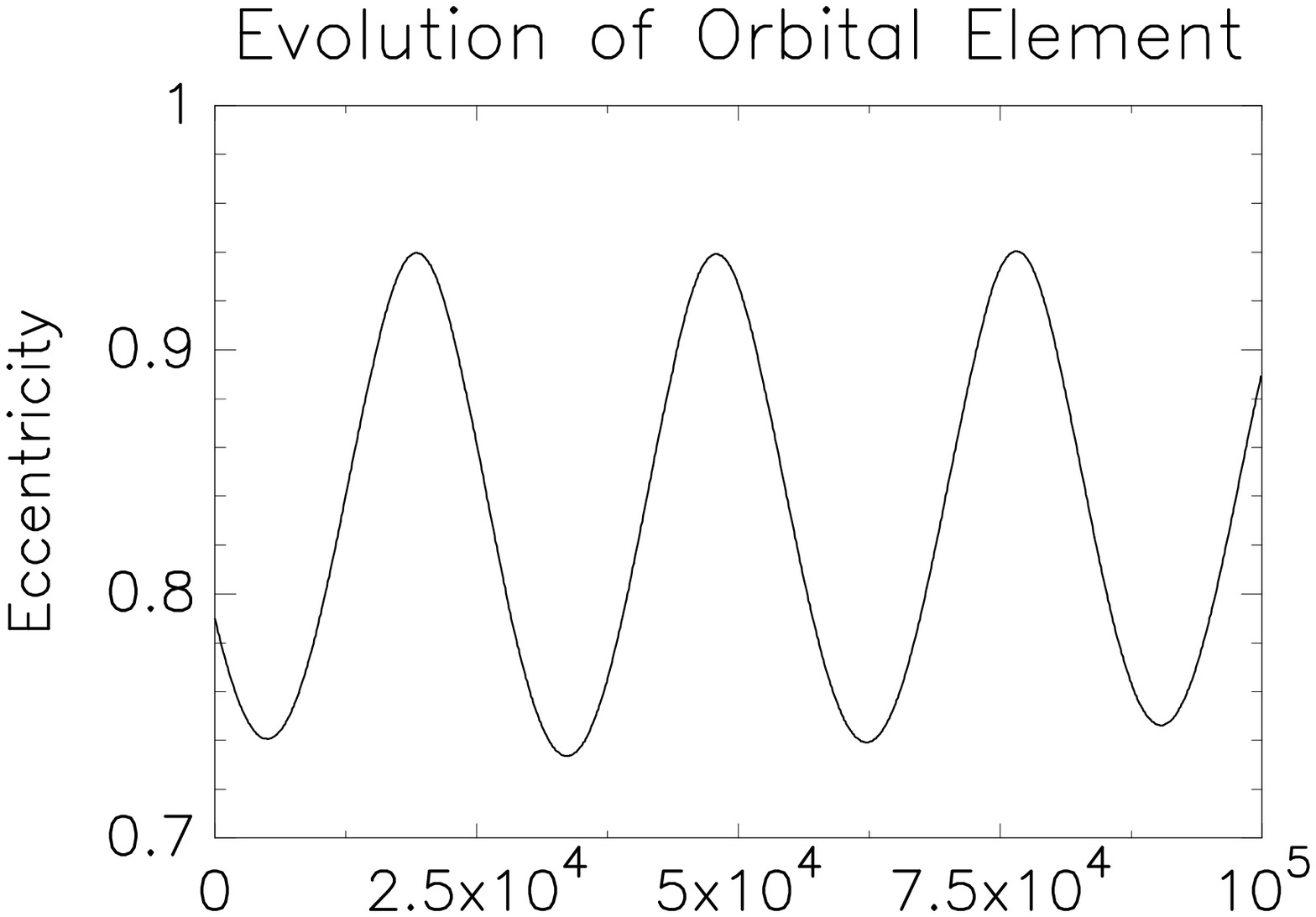}
\\[-4mm]
(b)\\[-\baselineskip]
\includegraphics[width=\columnwidth]{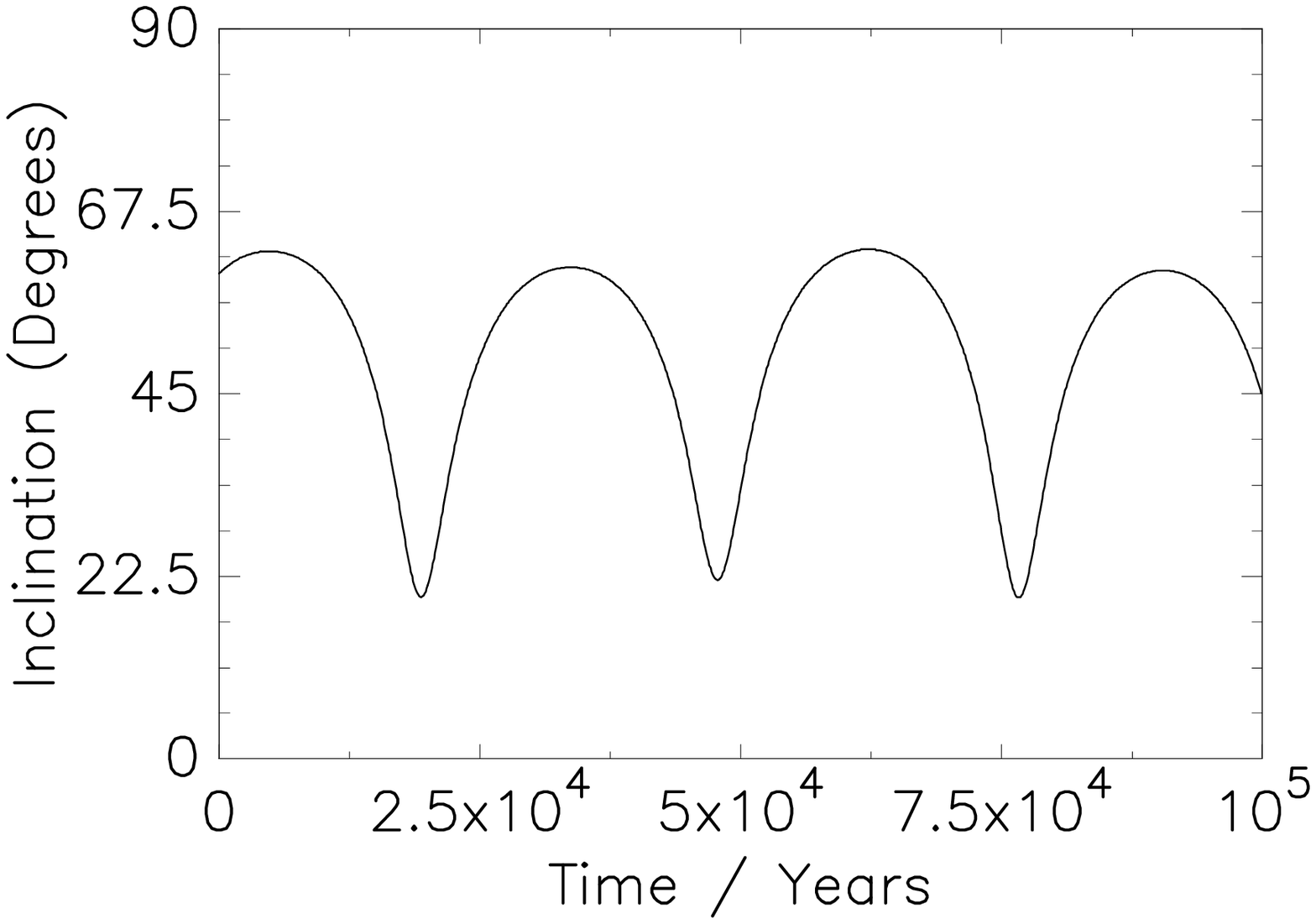}
\caption{Evolution in $e$ and $i$ of high $i$ asteroid 2008 KP. } 
\label{2008KP}
\end{figure}

To confirm our conclusions about 96P, we made ten clones by varying $a$ with $\pm\Delta a$ in equal steps where
$a$=3.03393972 and $\Delta a$=0.0001 au. Both Newtonian-only as well as GR-included n-body
numerical tests were conducted to check how both sets of bodies dynamically evolve
forward in time. 

\begin{figure}
(a)\\[-\baselineskip]
\includegraphics[width=\columnwidth]{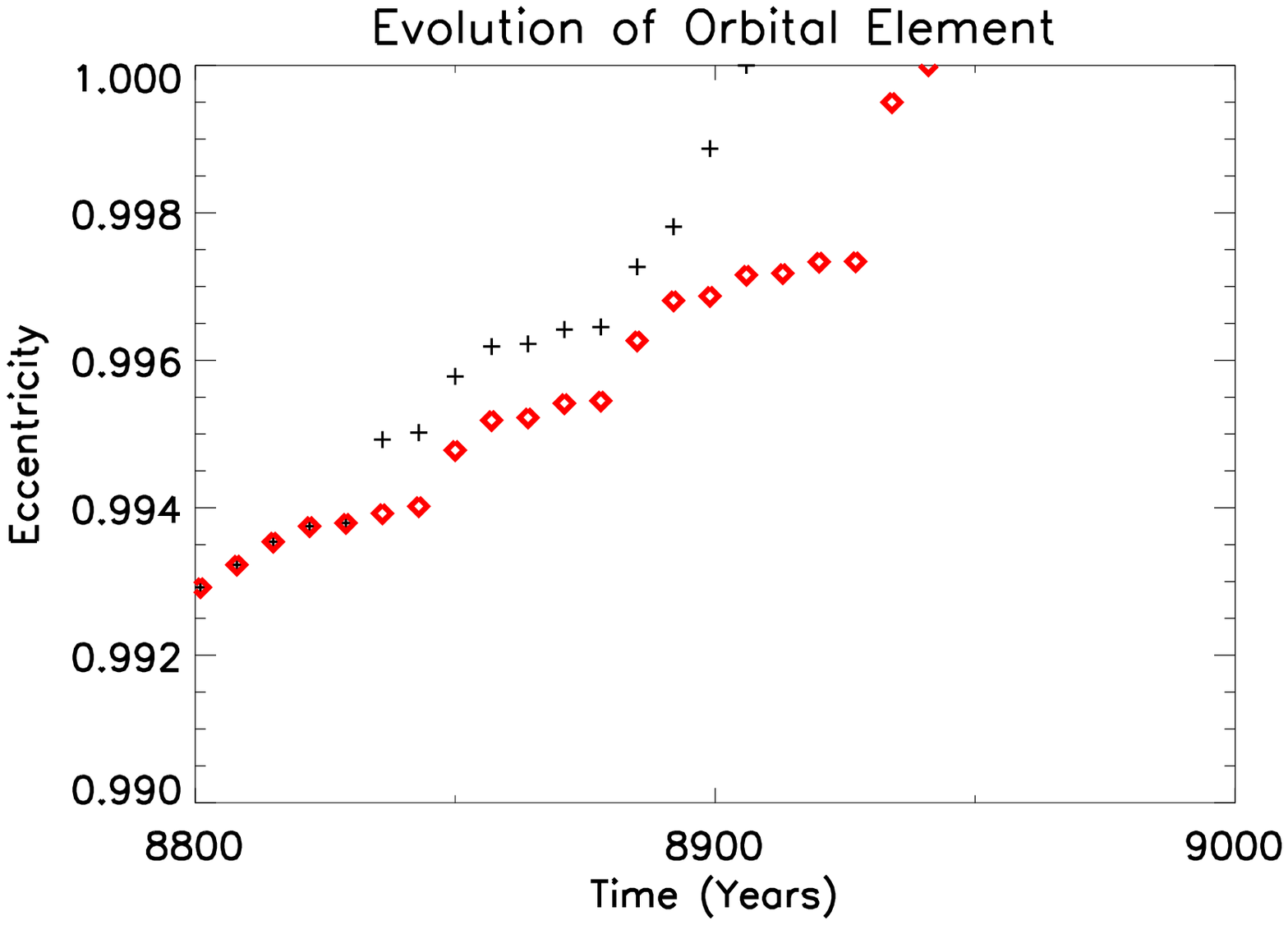}
\\[-5mm]
(b)\\[-\baselineskip]
\includegraphics[width=\columnwidth]{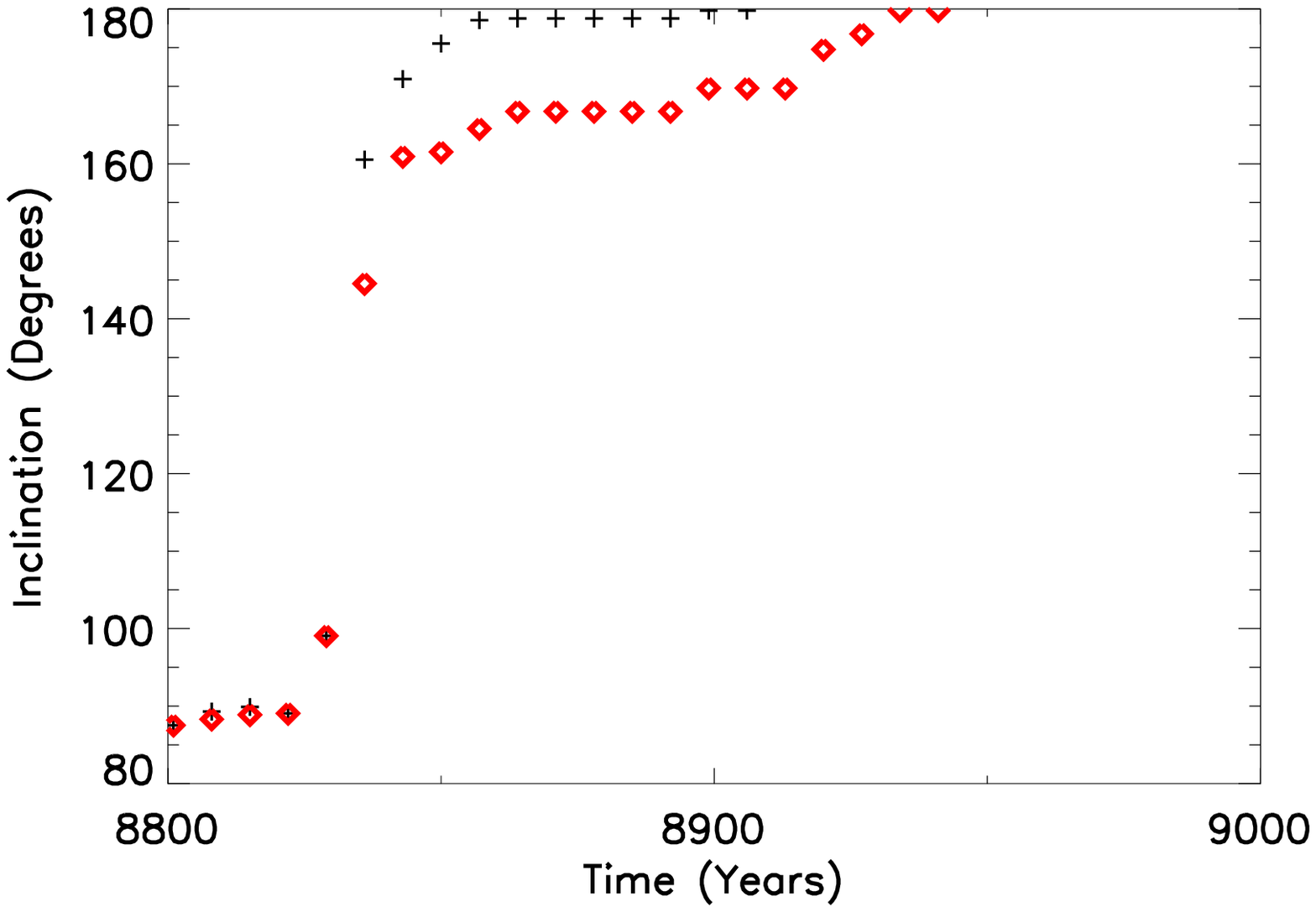}
\caption{Difference in dynamical behaviour of (a) eccentricity 
and (b) inclination between Newtonian-only (black cross) and
GR-included (red diamonds) cases, for final
$\sim$100 yr of 96P evolution. The particle falls into the sun earlier in
the Newtonian cases compared to the GR cases.}
\label{Flast100yr}
\end{figure}

The general dynamical behaviour of any 96P clone is practically identical in
the GR-included and Newtonian-only cases (Figure \ref{ei96P}), looking 
different only for the final sungrazing and sun
colliding phase (about 100 yr for 96P which is shown in
Figure \ref{Flast100yr}). The rate of change of $e$ and $i$ is different
towards the end of the bodies' evolution in the Newtonian-only and GR-included cases,
with $e$ reaching 1 later (typically 5--30 yr for different clones), and $i$
reaching 180\degr\ similarly later, in the GR case.  

\begin{figure}
\includegraphics[width=\columnwidth,trim = 15mm 90mm 20mm 80mm,clip]{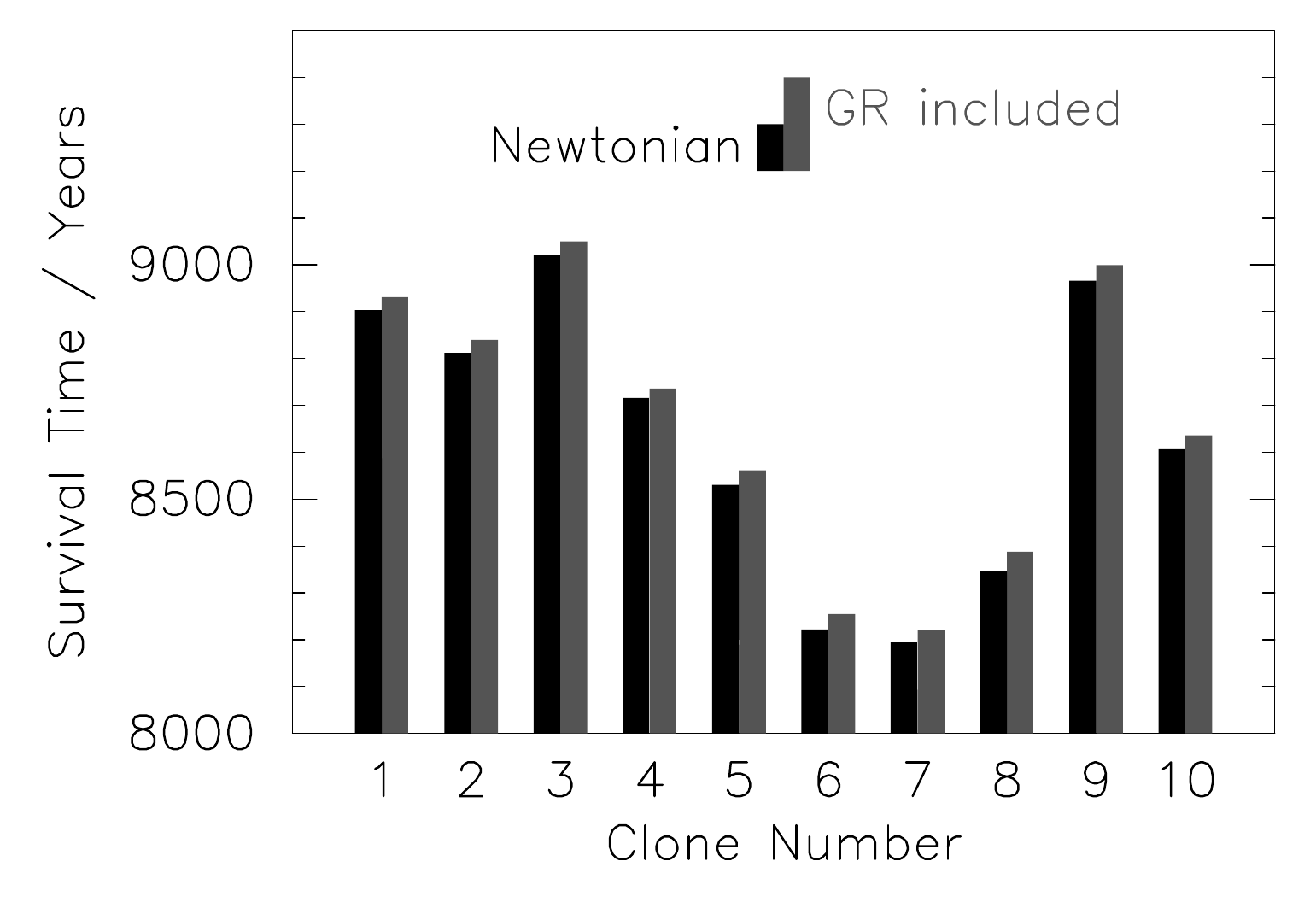}
\caption{Survival times of both sets of Newtonian-only and GR-included cases for all the 10 clones of 96P. The GR-included cases survive some extra years longer than Newtonian-only cases for all the clones in our integration.}
\label{SurvivalTime}
\end{figure}

The dynamical behaviour in forward integrations stays similar for both models
(which agrees with de la Fuente Marcos et al.\ 2015) in terms of inclination flips and
sungrazing plus sun colliding phases due to Lidov-Kozai like oscillations
occurring in both sets of integrations. We find that all the ten 96P clones
survive longer ($\sim$ 5--30 yr typically; shown in Figure \ref{SurvivalTime}) before colliding with the
sun in the GR-included integrations in comparison to the ten 96P clones in
Newtonian-only integrations. Figure \ref{SurvivalTime} shows that the ten clones survive similar timescales (varying by some hundred years) before falling into the sun. But at the same time, the difference in survival times between Newtonian-only and GR-included cases are consistent for every clone. 

De la Fuente Marcos et al.\ (2015) note that
retrograde phases in 96P are longer in GR-included simulations compared
to Newtonian-only simulations. In our simulations, the retrograde phase happens
only in the final $\sim$100 yr of 96P (during the same Lidov-Kozai like cycle
for all clones) and we do not see any notable change in its duration between
Newtonian-only and GR-included integrations.

Our integrations of 96P clones give an indication to the likely timeframes
when possible inclination flips and sun colliding episodes could occur in
future. We find that the timeframes are different from previous studies
(Gonczi et al. 1992, Levison \& Dones 2014, de la Fuente Marcos et al.\ 2015)
focused on 96P. This could be due to the relatively strong (compared to
strength of close approaches with other planets like Venus, Earth, Mars and
Jupiter, which we see in our integrations) close encounters with Mercury
during the initial phase of our integration. The different initial conditions and
orbital uncertainties have led to differences in results in previous works as well; this point has been
mentioned in de la Fuente Marcos et al.\ (2015). Hence it is evident
that exact inclination flip and sun colliding timings depend on the exact
initial conditions, the type of algorithm/integrator and number of gravitational perturbers (which is different
in the case of de la Fuente Marcos et al.\ 2015 compared to previous works) considered in the
integrations. In our work, initial conditions for 96P were taken from JPL-Horizons for epoch JD 2456541.5 (which is same as in de la Fuente Marcos et al.\ 2015) but the particles were integrated including the gravitational effects of eight planets only (whereas
de la Fuente Marcos et al.\ 2015 included the effects of dwarf planet
Pluto-Charon system and the 10 most massive main belt asteroids; their
initial conditions were taken in the barycentric frame from JPL-Horizons for
JD 2457000.5 and used a Hermite integrator for their calculations). In our
work, initial conditions for eight planets were taken in the heliocentric frame
for JD 2451000.5 and we employed Mixed-Variable Symplectic or RADAU
algorithms. Hence the difference in results is not surprising.  

Having said this, it should be noted that irrespective of the
small changes in initial conditions, all these clones undergo inclination
flip and sungrazing plus sun colliding phases at some point in time
(different clones showing change in times of the order of few hundreds of
years typically; Figure \ref{SurvivalTime}), while staying within the orbital phase space where GR
effects are measurable. This is the central point relevant to this work. 

\begin{figure}
\includegraphics[width=\columnwidth]{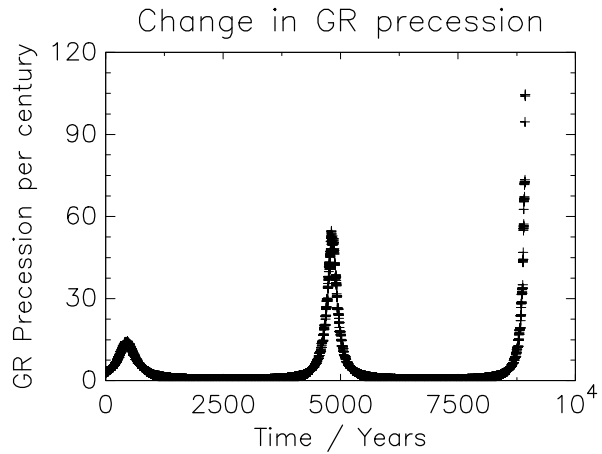}
\caption{Change in GR precession (in arc seconds) per century 
for 96P/Machholz 1 for 9 kyr from present. For comparison, GR precession of Mercury is 43 arc seconds per century.} 
\label{PrecCentury}
\end{figure}

\begin{figure}
\includegraphics[width=\columnwidth]{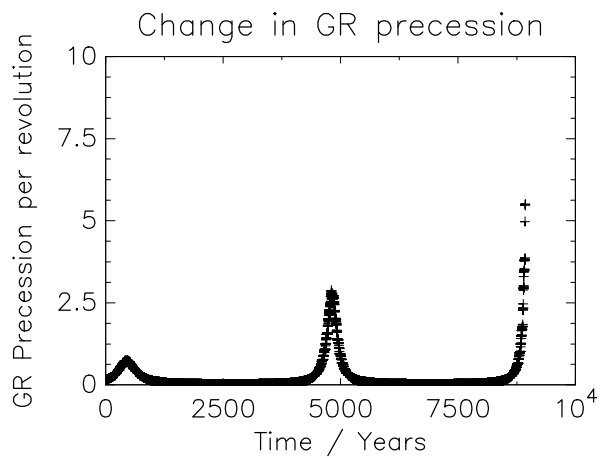}
\caption{Change in GR precession (in arc seconds) per revolution for
96P/Machholz 1 for 9 kyr from present. Drastic changes in GR precession rate
occur during some intervals. For comparison, GR precession of Mercury is
0.104 arcsec/rev.} 
\label{PrecRevolution}
\end{figure}

\begin{figure}
\includegraphics[width=\columnwidth]{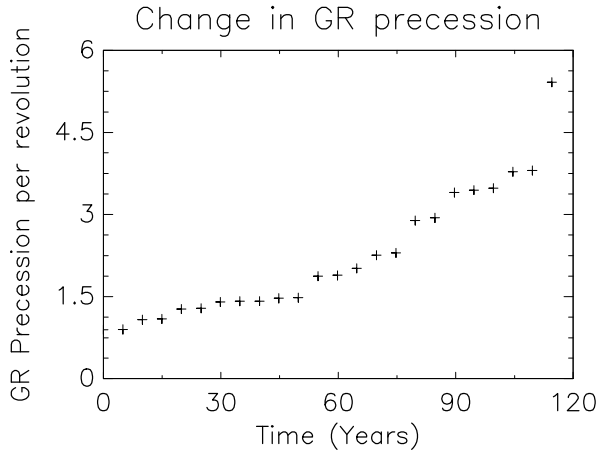}
\caption{Change in GR precession (arcsec/rev) for 96P/Machholz 1 for about
final 120 yr of same clone shown in Figure 8. There is a steady increase in
GR precession rate during the final phase of a sungrazing and sun colliding
orbit.}
\label{PrecShortTerm}
\end{figure}

Because of sungrazing behaviour during the final phase of 96P's evolution, GR
precession per revolution drastically increases in comparison to the initial
phase of its evolution. Figures \ref{PrecCentury} and \ref{PrecRevolution} show the GR precession per century
and per revolution respectively; $P$ stays approximately constant. Figure \ref{PrecShortTerm}
shows the change in GR precession for each revolution ($\sim$ 5 yr) in the
final 100 years.
These Figures are for clone 1 in Figure \ref{SurvivalTime}.

In Figure \ref{PrecRevolution}, there are three intervals where GR precession rate peaks to a
maximum. During 410--494 yr, GR precession $\sim$ 0.7 arc seconds per
revolution ($\sim$ 7 times that of Mercury's GR precession). During
4650--4978 yr, GR precession stays $\geq$ 1.0 arcsec/rev ($\sim 10 \times$
Mercury) with peak value $> 3$ arcsec/rev. During 8821--8926 yr, GR
precession again stays $\geq$ 1.0 arcsec/rev, peak value reaching about 6
arcsec/rev ($\sim 60 \times$ Mercury). These extreme points in GR precession
rate are due to direct effects of Lidov-Kozai like mechanism. 

\begin{figure}
\includegraphics[width=\columnwidth]{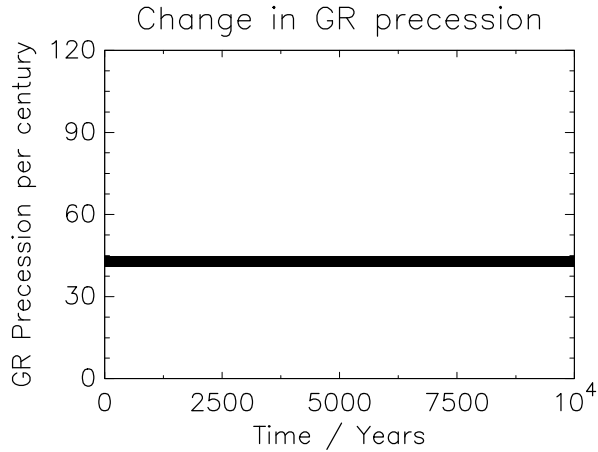}
\caption{Planet Mercury's GR precession (arcsec/century)
for 10 kyr from present. }
\label{PrecMercury}
\end{figure}

Our simulations show that the combination of Lidov-Kozai like oscillations and GR
precession lead to these striking sudden changes in the rate of GR precession
at different points in time. In contrast with 96P, the change in GR
precession rate of Mercury itself (see Figure \ref{PrecMercury}) remains almost constant
throughout the same period. These are two extreme situations in the solar
system in terms of this behaviour related to change in GR precession rates. 

We find that Lidov-Kozai like oscillations, evolution of Keplerian orbital elements
and the general dynamical behaviour remain the same in Newtonian and GR models
for the bodies discussed in this work. However we were particularly interested in the rate of change of GR precession
for individual bodies due to rapid changes in perihelion distances induced by
Kozai oscillations and this particular pattern (due to combination of GR and
Kozai) is the highlight of this study.

When Lidov-Kozai like oscillations and GR precession co-exist for the same
particle, the secular decrease in $q$ caused by the former can lead to
sungrazing and sun colliding phases which in turn can lead to drastically
higher GR precession than at the starting time, during the final phase of
these orbits (see Figures \ref{PrecCentury}, \ref{PrecRevolution} and \ref{PrecShortTerm}) before they fall into the sun. This
phenomenon becomes even more dynamically interesting when inclination flips
occur during the final sungrazing or sun colliding phase for a body. Such a
unique example in the solar system is provided by comet 96P. For the 96P
clone having the highest contrast in GR precession rates during its evolution
(see Figures \ref{PrecCentury} and \ref{PrecRevolution}), Table \ref{tab-96P} lists data for the first and last
100 yr. For other timeframes in between, GR precession rates also increase
and decrease rapidly, whereas for bodies like Mercury (or other terrestrial
planets) GR precession per revolution stays roughly constant (Figure \ref{PrecMercury}) for
a long time. During the final sungrazing and sun colliding phase the GR
precession for the final 100 yr in 96P exceeds that of Mercury's GR
precession per century. Moreover although GR precession increases
substantially from 96P's initial to final sungrazing phase, conversely the
Newtonian precession substantially reduces between these two phases
(Table \ref{tab-96P}). Respective values for Mercury are given in the
Table \ref{tab-96P} caption for comparison. 

The disparity between analytical and numerical values of GR precession during
the final sungrazing phase in Table \ref{tab-96P} is because of the rapid
change (in 100 yr) in the body's orbit which reduces $q$ rapidly and hence is
not taken into account when computing values analytically using Equation
1. On the other hand, in numerical integrations the rapid changes in $q$
values at different time steps are correctly accounted for while calculating
GR precession and hence this value is more reliable in the final sungrazing
phase. Although the magnitude of Newtonian precession always exceeds GR
precession at any instant of time, the Newtonian and GR precession display
opposite trends during the sungrazing phase (Table \ref{tab-96P}). This
behaviour is typical for all the 96P clones in our simulations.

\begin{table}
\caption{Change in GR and Newtonian precession in argument of pericentre per
century ($\Delta \omega/\Delta t = \omega_{t_{2}}/t_{2}- \omega_{t_{1}}/t_{1}$)
for a typical clone of 96P during initial and final hundred years of its
evolution. For comparison, the GR and Newtonian precession values for Mercury
are 43 and 5557 arc seconds/century (page 199, Weinberg 1972). Initial
orbital elements are from JPL-Horizons and final elements from two
independent numerical integrations (Newtonian-only and GR-included) performed
using MERCURY package to compute the differences in $\omega$ which gives the
numerical GR and Newtonian precession. $a \sim 3.034$ au during both initial
and final phases of dynamical evolution discussed here.}
\vspace{0mm}
\begin{center}
\begin{tabular}{@{}l@{\,\,\,\,}l|l@{\,\,\,\,}l|l@{}} \hline
Body & $q$ &$\Delta \omega/\Delta t$ & $\Delta \omega/\Delta t$  & $\Delta \omega/\Delta t$ \\
& (au)               & (arcsec/      & (arcseconds  & (arcseconds \\
&                    &  century)     &     per century)    & per century)   \\
&                        & (analytic)   &  (integrations)       & (integrations) \\
&                        & (GR)   &  (GR)       & (Newtonian) \\
\hline \hline
96P/Machholz 1  & 0.124  & \phantom{0}3.0 & \phantom{0}3.6  & 3533.0 \\
(for first 100 yr  &                      &                   &  &     \\
in the simulation)  &                      &                   &  &     \\
96P/Machholz 1  & 0.022            & 16.9 & 53.3  & \phantom{0}285.2 \\
(for last 100 yr  &                     &                   &  &     \\
in the simulation)  &                     &                   &  &     \\
\hline
\end{tabular}
\end{center}
\label{tab-96P}
\end{table}

Although some areas of the solar system can be extremely chaotic in
general, the test particles evolving in the gravitational models we present
here appear to follow a reasonably predictable pattern, that is, the
evolution of orbital elements into the future is predictable to a good degree
as a function of initial epoch and initial orbital elements. This is
supported by the different evolution of clones with initial separation
$\Delta a = 0.0001$ au (survival times in Figure \ref{SurvivalTime} varying
up to several hundred yr), whereas with smaller $\Delta a = 10^{-6}$ au (not
plotted here) then the clones have practically identical dynamical
evolution.

\section{Conclusion and Future Work} 

We have shown that there are bodies in the solar system in which both GR
precession and Lidov-Kozai like mechanism can co-exist and be comparable in their effects and for
which these complementary effects can be measured and identified using analytical and
numerical techniques. Thus there is a continuum of bodies encompassing,
firstly GR precession dominant, secondly GR precession plus Lidov-Kozai like
mechanism co-existing and finally Lidov-Kozai like mechanism dominant states which
are all permissible in nature. A real solar system body in this intermediate
state is identified using compiled observational records from IAU-MPC,
Cometary Catalogue, IAU-MDC and performing analytical plus numerical tests on
them. This intermediate state brings up the interesting possibility of
drastic changes in GR precession rates (at some points peaking to about 60 times that of Mercury's GR precession) during orbital evolution due to
sungrazing and sun colliding phases induced by the Lidov-Kozai like mechanism,
thus combining both these important effects in a unique and dynamically
interesting way. Comet 96P/Machholz 1 stands out as the only real body
identified (from our simulations) to be exhibiting these interesting traits,
as well as inclination flips, in the near future. 

This study's purpose was to focus on the real small bodies in the solar
system (including the planetary perturbations during the present epoch)
exhibiting these two dynamical phenomena at the same time. For future work,
it would be instructive to do a detailed abstract study, with only Jupiter so
as to induce the Kozai mechanism (in its pure form) and exclude perturbations from other
planets, mapping the entire Keplerian elements phase space to find the
boundaries between three regions namely, GR precession dominant regime, GR
precession plus Kozai mechanism co-existing regime and Kozai mechanism
dominant regime. One could then compare the stability and chaotic levels
between these three regions. This would enable us to understand the various
patterns in change of GR precession rates and the maximum rates of GR
precession possible in a short time for bodies in our solar system due to
sungrazing phases driven by strong Lidov-Kozai like oscillations. Independently, a
study of this nature could tell us the exact phase spaces which can
contribute to possible sudden peaks in rates of GR precession due to Kozai like
oscillations in the context of stability of artificial satellite orbits (Rosengren et al. 2015) so that precise
measurements can be made for the confirmation of this combined phenomenon in
future long term satellites depending on the phase space traversed.

\vspace*{-6mm}

\section*{Acknowledgments} 
The authors thank Smadar Naoz for numerous valuable suggestions and improvements.
Sekhar and Werner acknowledge the Crater Clock project (235058/F20) based at Centre for
Earth Evolution and Dynamics (through the Centres of Excellence scheme
project number 223272 (CEED) funded by the Research Council of Norway) and
USIT UNINETT Sigma2 computational resource allocation through the Stallo cluster with project accounts
Notur NN9010K, NN9283K, NorStore NS9010K and NS9029K. Research at Armagh Observatory and Planetarium is funded by the
Department for Communities for N. Ireland. Vaubaillon thanks the CINES supercomputing facility of France. Li acknowledges Matt Payne for developing and providing the GR sub-routine for MERCURY.

\end{document}